\documentclass[%
 reprint,
 amsmath,amssymb,
 aps,
]{revtex4-2}
\pdfoutput=1
\usepackage{amsmath}
\usepackage{xcolor} 
\usepackage{graphicx}
\usepackage{dcolumn}
\usepackage{bm}
\usepackage{tabularx}
\usepackage{xurl}
\usepackage[breaklinks]{hyperref}

\begin{document}

\title{Design and simulation of a source of cold cadmium for atom interferometry}

\author{Satvika Bandarupally}
\author{Jonathan N. Tinsley}
\author{Mauro Chiarotti}
\altaffiliation[Also at ]{CNR-INO, Sesto Fiorentino, Italy}%
\author{Nicola Poli}
\altaffiliation[Also at ]{CNR-INO, Sesto Fiorentino, Italy}%
\email[Author to whom correspondence should be addressed: ]{nicola.poli@unifi.it}%
\affiliation{Dipartimento di Fisica e Astronomia and LENS, Universit\`a degli Studi di Firenze, Via G. Sansone 1, 50019 -- Sesto Fiorentino, Italy}

\date{\today}








\begin{abstract}
We present a novel optimised design for a source of cold atomic cadmium, compatible with continuous operation and potentially quantum degenerate gas production. The design is based on spatially segmenting the first and second-stages of cooling with the the strong dipole-allowed $^1$S$_0$-$^1$P$_1$ transition at 229~nm and the 326~nm $^1$S$_0$-$^3$P$_1$ intercombination transition, respectively. Cooling at 229~nm operates on an effusive atomic beam and takes the form of a compact Zeeman slower ($\sim$5~cm) and two-dimensional magneto-optical trap (MOT), both based on permanent magnets. This design allows for reduced interaction time with the photoionising 229~nm photons and produces a slow beam of atoms that can be directly loaded into a three-dimensional MOT using the intercombination transition. The efficiency of the above process is estimated across a broad range of experimentally feasible parameters via use of a Monte Carlo simulation, with loading rates up to 10$^8$~atoms/s into the 326~nm MOT possible with the oven at only 100~$^\circ$C. The prospects for further cooling in a far-off-resonance optical-dipole trap and atomic launching in a moving optical lattice are also analysed, especially with reference to the deployment in a proposed dual-species cadmium-strontium atom interferometer. 
\end{abstract}

\maketitle


\section{Introduction} \label{sec:intro}

The production of cold, large and dense samples is an indispensable technique in modern atomic, ionic and molecular physics~\cite{Schreck_2021}. It forms the experimental basis of a diverse range of fundamental and applied experiments, including frequency metrology~\cite{Poli_2013}, searches for exotic matter and forces~\cite{Safronova_2018}, and atom interferometry~\cite{Varenna_2014}. Techniques for the fast and robust generation of ultracold samples of many species are consequently well established, for example in the atomic domain, Cs, Rb, Sr, Yb and many others. In other cases, source preparation remains difficult, especially for molecules, due to complexities of the energy level structure or the availability of suitable lasers. In particular, there is growing interest in the laser cooling and trapping of alkaline-earth-like metals, such as Cd, Zn and Hg~\cite{Yamaguchi_2019,Zhang_2021,Buki_2021,Lavigne_2022}, whose deployment has been slowed by the relevant cooling and trapping transitions lying in the challenging ultraviolet regime.

\begin{figure}[t]
\centering\includegraphics[width=0.5\textwidth]{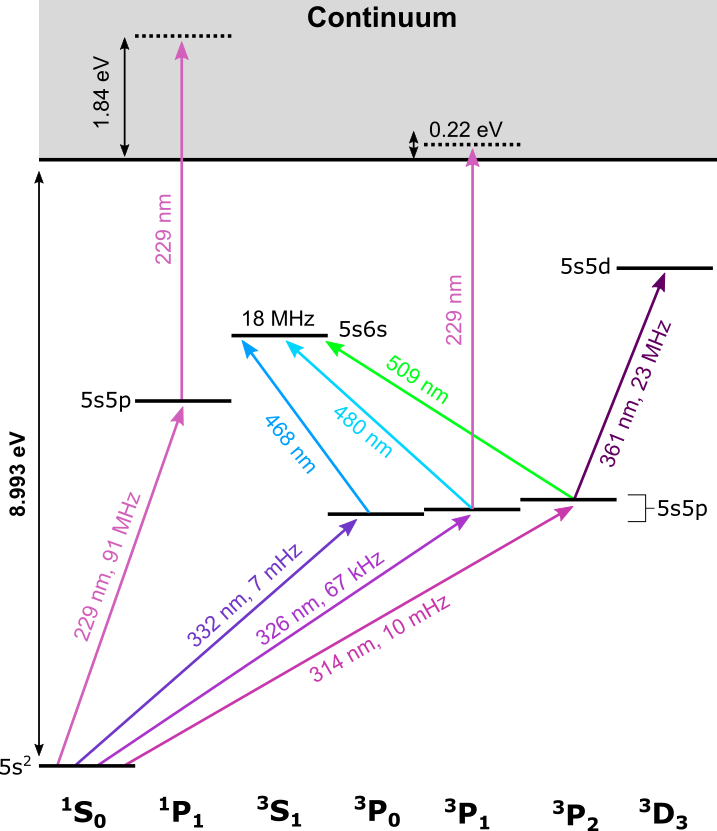}
\caption[Energy level diagram of cadmium with main transitions]{(a) To-scale partial energy level diagram and main transitions for bosonic Cd. These levels and transitions shown have previously been used for cooling and trapping Cd. Also shown is the photoionsiation energy and the energy of a 229-nm photon starting from the important $^1$P$_1$ and $^3$P$_1$ states.}
\label{fig:transitions}
\end{figure}

Here we focus on the design and simulation of a high-flux source of atomic cadmium, which is a transition metal possessing two valence-shell electrons and a similar transition structure to alkaline-earth atoms, providing access to narrow-linewidth intercombination transitions, ideal for high-precision metrology, such as optical clocks
and atom interferometers~\cite{Tinsley_2022}, and also access to a broad, dipole-allowed transition suitable for rapid cooling of room-temperature atoms to the mK regime (Fig.~\ref{fig:transitions}). In comparison to other alkaline-earth and alkaline-earth-like systems, e.g. Sr and Yb, which have been extensively utilised in leading optical clocks~\cite{Takamoto_2005,McGrew_2018,Bothwell_2022,Zheng_2022} and which are being utilised in a raft of next-generation interferometers~\cite{Hu_2017,AION_2020}, in Cd these transitions lie in the UV region, enhancing intrinsic measurement sensitivity of clocks and interferometers and dramatically reducing the sensitivity to blackbody radiation, a major systematic error in clocks~\cite{Itano_1982,Bothwell_2019} and also a factor in high-precision atom interferometry~\cite{Haslinger_2018}. The afforded high scattering rate and low wavelength of $^1$S$_0$-$^1$P$_1$ dipole-allowed transitions may potentially also benefit single-atom optical tweezer experiments and related quantum simulators~\cite{Kaufman_2021}.

Despite the increasing interest in Cd due to these properties, experimental demonstrations of cold Cd available in the literature are limited to a handful of examples, with demonstrations of magneto-optical traps (MOTs) on the broadband $^1$S$_0$-$^1$P$_1$ transition at 229~nm~\cite{Brickman_2007,Kaneda_2016} and, more recently, on the narrow 326~nm $^1$S$_0$-$^3$P$_1$ transition~\cite{Yamaguchi_2019}. Other common techniques such as Zeeman slowers and 2D-MOTs, or the use of spatially separated regions for optimal vacuum pressure levels, have not been reported, even though they form the basis of many experiments optimised for fast atom loading or continuous sources~\cite{Chen_2019}. Similarly, the production of quantum degenerate sources of Cd has yet to be reported. All attempts to cool and trap Cd are hampered by the problematic nature of the 229~nm light, which is difficult to produce stably at high continuous-wave powers, damages vacuum components and causes photoionisation of Cd~\cite{Brickman_2007}. Very recently, a system to trap atoms without using the $^1$S$_0$-$^1$P$_1$ transition light has been reported, with atom numbers in an intercombination transition MOT enhanced by using the 23-MHz-wide $^3$P$_2$-$^3$D$_3$ transition at 361~nm and two further lasers for optical pumping~\cite{Schussheim2018,Ohayon_2022}.

\begin{figure}[t]
\centering\includegraphics[width=0.45\textwidth]{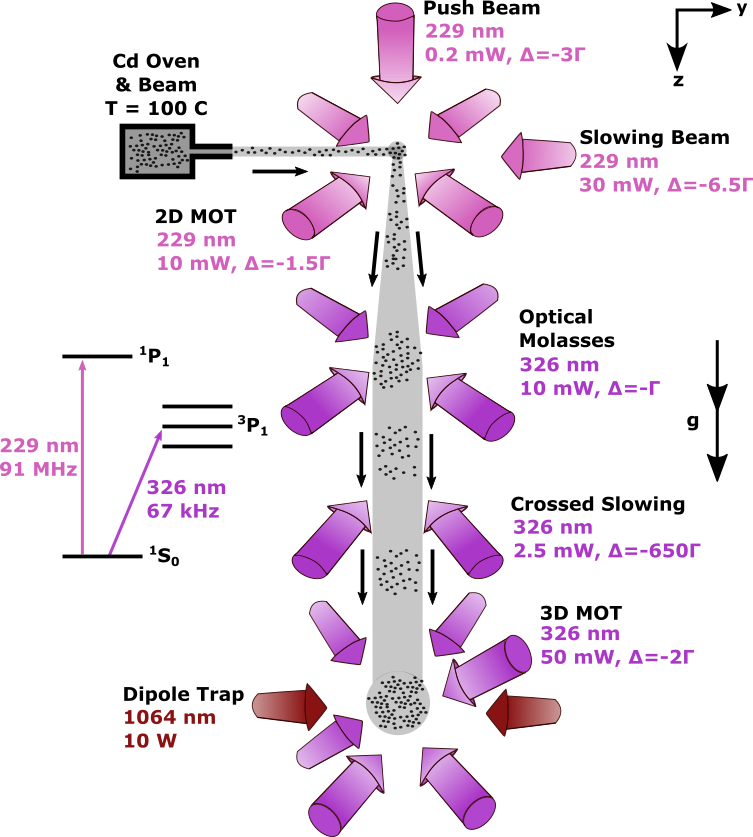}
\caption[Basic idea of the cold Cd Source]{Cartoon of the basic idea of the cold Cd source, using the $^1$S$_0$-$^1$P$_1$ transition at 229~nm and $^1$S$_0$-$^3$P$_1$ transition at 326~nm in spatially separated regions. The approximate required power per beam and frequency detuning of each stage is shown.}
\label{fig:idea}
\end{figure}

In this article, we present the design and simulation of such an optimised system for Cd to be used as the atomic source for an atom interferometer~\cite{Tinsley_2022}, and with a focus on the unique challenges and opportunities of this atom. In particular, we have designed and extensively simulated a system which uses only a minimal amount of 229~nm light to generate a slow beam of atoms which can be trapped directly and efficiently in a MOT based only on the 326~nm intercombination transition, the basic idea of which is shown in Fig.~\ref{fig:idea}. The design is inspired by recent developments in continuous source production~\cite{Lamporesi_2013,Bennetts_2017} and is centred around two recently demonstrated UV laser systems~\cite{Tinsley_2021,Manzoor_2022}, which have been designed specifically for this purpose, and on a novel effusive atomic beam of Cd~\cite{Tinsley_2021}. 

The structure of this article is the following; in Section~\ref{sec:coldCd} we discuss the general design requirements for the cold-atom source apparatus and discuss the relevant properties of Cd in detail; an overview and the basic idea of the source is given in Section~\ref{sec:overview}; in Section~\ref{sec:simulation} we present the atom-light interaction model used and give details of the numerical simulation; simulation results of the first two-stages of cooling at 229~nm are presented in Section~\ref{sec:2dmot}; and likewise the trapping in a 3D MOT at 326~nm is shown in Section~\ref{sec:3dmot}; Section~\ref{sec:vacuum} brings these results together to present a finalised vacuum chamber system and a numerical simulation of the full system; Section~\ref{sec:odt} presents the design of optical-dipole-trap systems for the further cooling, spatial transfer and launching of the atoms; finally, conclusions and the experimental outlook are reported in Section~\ref{sec:summary}.

\section{Cadmium Characteristics \& Atomic Source Requirements}\label{sec:coldCd}

An ideal cold atom apparatus for quantum experiments should be able to load large numbers of atoms in a vacuum chamber where background gas collisions are negligible over the timescales of the experiment to preserve coherence~\cite{Ferrari_2006}. Large atom numbers are required to minimise the quantum projection noise (or standard noise limit)~\cite{Itano_1993}, which is often the limiting sensitivity factor for atom interferometers~\cite{Sorrentino_2014}. Moreover, the cold atom preparation should be as rapid as possible to enhance sensitivity~\cite{Savoie_2018} and minimise frequency aliasing problems arising from the Dick effect~\cite{Quessada_2003}.
Practically, these requirements often require a high-flux source of atoms which can be efficiently trapped in a science chamber, which is spatially segmented from the source. 
Finally, we note that the system should be robust, allowing for stable operation over the months and years typically required to perform cold atom experiments.

A simplified energy level diagram with allowed transitions for bosonic cadmium is shown in Fig.~\ref{fig:transitions} and further details on the key $^1$S$_0$-$^1$P$_1$ and $^1$S$_0$-$^3$P$_1$ transitions for laser cooling and trapping are given in Table~\ref{tab:atomProperties}, with the values for the same transitions in the more commonly employed Sr and Yb also provided for comparison. In this section, we will discuss these two transitions in detail and highlight how their features guide our cooling and trapping apparatus design, also considering the requirements highlighted above.

\begin{table*}
 \caption{Relevant properties of the first and second-stage cooling transitions for Cd, Sr and Yb. Values for the wavelength ($\lambda$), corresponding natural linewidth ($\Gamma$), saturation intensity ($I_s$) and the Doppler temperature ($T_D$) are reported. Also shown is the stopping distance ($d_{\text{stop}}$) for atoms at the most probable velocity of a beam ($v_b=\sqrt{3k_BT/m}$), where the beam temperature ($T_b$) is set to give approximately the same vapour pressure as Cd at 100~$^\circ$C (2.5$\times$10$^{-7}$~mbar).} 
 \label{tab:atomProperties}
 \begin{center} 
 \begin{tabular}{l c c c c c c c}
    \hline
    \hline
    Atom          &$\lambda$ (nm)    &  $\Gamma/2\pi$         & $I_s$ (mW/cm$^2$)     & $T_D$ & $T_b$ ($^\circ$C ) & $v_b$ (m/s) & $d_{\text{stop}}$ (m)\\
    \hline
    \multicolumn{8}{c}{$^1$S$_0$-$^1$P$_1$ (First-stage)}\\
    \hline
       Cd &   228.9         &  91~MHz                & 991                & 2.2~mK & 100 & 288 & 9$\times$10$^{-3}$\\
           Sr &   460.9         &  32~MHz                 & 42.5               & 0.8~mK & 280 & 397 & 8$\times$10$^{-2}$\\  
       Yb &   398.9         &  28~MHz             & 57.7               & 0.7~mK & 240 & 272 & 7$\times$10$^{-2}$\\
    \hline 
    \multicolumn{8}{c}{$^1$S$_0$-$^3$P$_1$ (Second-stage)}\\
    \hline
       Cd &   326.1          &  66.6~kHz                & 0.252            & 1.6~$\mu$K & 100 & 288 & 18\\
       Sr &   689.4         &  7.4~kHz                & 0.003   & 180~nK & 280 & 397 & 513\\
       Yb &   555.8         &  182.2~kHz                & 0.139   & 4.4~$\mu$K& 240 & 272 & 16\\
    \hline     
    \hline
 \end{tabular}
 \end{center}
\end{table*}

The dipole-allowed transition $^1$S$_0$-$^1$P$_1$ has two very noticeable features: a short wavelength of 229~nm lying in the deep-ultraviolet (DUV) regime; and a very broad natural linewidth (2$\pi\times$91~MHz)~\cite{Xu_2004}. Both these features present technical challenges in implementing an optimised practical realisation of a cold Cd atom source. Ideally, the system would be operated close to the saturation intensity to saturate the cooling and trapping forces, but given the very high saturation intensity of the main broad cooling transition (991~mW/cm$^2$), this would require high continuous-wave powers at 229~nm which is problematic for a number of technical and fundamental reasons.

Firstly, the laser sources and the optics in the DUV regime are still under progress and are relatively less developed in comparison to visible or infrared regime. Although UV lasers are constantly improving, the regime below approximately 240~nm remains highly challenging. For example, recent advances using CLBO crystals to generate 2~W of stable power at 261~nm~\cite{Shaw_2021} cannot be directly applied due to the phase-matching properties of CLBO, which has a Type-I SHG cut-off wavelength at $\sim$237~nm~\cite{Bhar_2000}. Instead beta-barium borate (BBO) crystals have to be used, which exhibit greater DUV-induced damage~\cite{Takachiho_2014}, and achieving such powers stably over significant timescales ($>$hours) has not currently been demonstrated. Additionally, many optical components are not readily available at this wavelength. For example, while single-mode fibres for the UV have been demonstrated, they are not commercially available and require a complex production procedure for a still limited performance~\cite{Gebert_2014,Marciniak_2017}.

Moreover, even if W-level power at 229~nm were available, it is not clear that it would be advantageous for a practical experiment with Cd atoms. One issue is that at this short wavelength, Cd atoms are prone to photoionisation from both the $^1$P$_1$ and, potentially, the $^3$P$_1$ states (Fig.~\ref{fig:transitions}), which will practically limit the number of atoms which can be loaded into a MOT~\cite{Brickman_2007}. For example, the expected loss rate due to photoionisation can be calculated according to $\Gamma_{\text{ion}}=\sigma If/\hbar\omega$, where $I$ is the beam intensity, $f$ is the fraction of atoms in the $^1$P$_1$ state and $\sigma$=~$2\times10^{-16}$~cm$^2$~\cite{Brickman_2007}, meaning that for a MOT using 6 beams with $P$=150~mW, $w$=2~mm ($I$=2400~mW/cm$^2$, $s$=2.4) and detuning $\Delta$=$-\Gamma$, we obtain a photoionisation loss rate $\Gamma_{\text{ion}}=$1.2~kHz which represents a significant loss for the MOT. Furthermore, high-power DUV light is damaging to optical coatings, especially those under vacuum, leading to degradation of performance. The complete damage mechanisms are not fully understood, though there seems to be contributions from both oxygen depletion of the coating material and from various UV-induced mechanisms related to hydrocarbon contamination~\cite{Hollenshead_2006,Gangloff_2015}. Although research into improving optical coatings under vacuum is ongoing~\cite{Gangloff_2015,Hubka_2021} and fluoride-based coatings seem to perform much better~\cite{Burkley_2021}, this problem is not currently solved. Excessive use of 229-nm light would therefore require the system to be regularly opened or purged to recover the coating performance.

Also mentioned in Table~\ref{tab:atomProperties} is the large linewidth of the 229-nm transition. For the development of the MOT on this transition with high scattering rate, high magnetic field gradients are prescribed. For example, a MOT can be modelled as a damped harmonic oscillator~\cite{Steane_1992} and we estimate that critical damping requires a gradient of $\mathrm{d}B/\mathrm{d}z\sim210$~G/cm for saturation intensity $s$=0.2 and detuning $\Delta$=$-\Gamma$. 
Such a requirement would rule out the usage of water-cooled magnetic coils not only due to the sheer blockage of optical access, but also the possibility of eddy currents arising from  switching the large currents ($\sim$100~A). 

However, this combination of the short wavelength and broad natural linewidth does prevent significant advantages if utilised correctly. 
It allows for a very large deceleration force on the atom which can, for example, dramatically shorten the length of the Zeeman slower stage where the atoms from the room temperature are cooled down to tens of m/s over a distance of cm. For example, the computed minimum stopping distance for atoms travelling at a speed $v$=290~m/s is only 9~mm (Table~\ref{tab:atomProperties}). Furthermore, the broad transition means that the approximate capture velocity of any MOT can potentially be large at practical values ($>$50~m/s), making it easy to capture atoms from e.g. a relatively fast atomic beam.

The Doppler temperature of the $^1$S$_0$-$^1$P$_1$ is, however, higher than desirable at 2.2~mK, so in any case further cooling is mandatory. Alkaline-earth and alkaline-earth-like systems typically achieve cooling to the necessary $\mu$K regime by using a second-stage MOT on the narrow $^1$S$_0$-$^3$P$_1$ intercombination transition, with the laser frequency modulated to enhance its scattering force~\cite{Kuwamoto_1999}. Although this transition is also in the UV regime for Cd at 326~nm, it is considerably less challenging and powers approaching the W-level around this wavelength can be more readily achieved~\cite{Hannig_2018,Manzoor_2022,Chiarotti_2022}.

One intriguing possibility is the direct loading into the intercombination-transition MOT of an atomic beam of Cd, something which is routinely performed for Yb with loading rates of 10$^8$~atoms/s~\cite{Guttridge_2016}, but is far more challenging for Sr, though possible in a carefully optimised system~\cite{Bennetts_2017}. As shown in Table~\ref{tab:atomProperties} the linewidth of the $^1$S$_0$-$^3$P$_1$ transition of Cd lies between these two cases, suggesting a difficult but achievable process, as also suggested by the similarity in the stopping distance of Cd and Yb (Table~\ref{tab:atomProperties}). The drawback of this technique is that the capture velocity of the MOT will be limited to $\sim$5~m/s, so only a very small fraction of room-temperature atoms can captured if used on its own.

\section{Overview \& Idea of the system}\label{sec:overview}
The basic idea of system is shown in Fig.~\ref{fig:idea}, inspired by previous systems with other species~\cite{Lamporesi_2013,Bennetts_2017}. In brief, the main guiding principle was to minimise the amount of 229~nm light required and to separate the cooling based on the broad and narrow transitions at 229~nm and 326~nm, respectively, principally due to the aggressive and problematic nature of 229~nm light discussed above (see Section~\ref{sec:coldCd})

Cadmium atoms will be emitted from an oven at 100~$^\circ$C to form an effusive atomic beam which is loaded into a 2D MOT on the broad $^1$S$_0$-$^1$P$_1$ transition at 229~nm. Unlike a 3D MOT, this system does not have steady-state trapping of the Cd atoms and therefore reduces the interaction time of the atoms with the 229~nm light. The magnetic field for a 2D MOT can furthermore be generated with a simple arrangement of permanent magnets~\cite{Lamporesi_2013}, removing the requirement of high electrical currents otherwise necessary for generating of the required large gradients ($>$200~G/cm). The loading rate of this 2D MOT can be optionally enhanced by using a transverse-field Zeeman slowing beam, without the need for further magnets. 

A low-intensity push beam will both plug the 2D MOT in the non-trapping dimension and direct the atoms from the 2D MOT vertically downwards towards a chamber, around 35~cm beneath the 2D MOT, where they will be loaded directly into a 3D MOT based on the narrow $^1$S$_0$-$^3$P$_1$ 326~nm transition. Separating these two MOTs spatially, rather than temporally, is beneficial for improving vacuum quality and potentially allows for a continuous flux of cold atoms~\cite{Bennetts_2017,Chen_2019}. In the specific case of Cd, there are additional practical benefits to separating the MOT regions; for example, this design lessens the problem of photoionisation by reducing the interaction time with the 229~nm light (see Section~\ref{sec:2dmot}) and protects the weak 3D MOT from the strong 229-nm photons.

For an atomic beam with a longitudinal velocity below the capture velocity of $\sim$5~m/s, the acceleration due to gravity begins to play a non-trivial role. For example, for an estimated transit distance of $\sim$30~cm, atoms with an initial longitudinal velocity in the horizontal direction of 4~m/s will fall 3~cm off axis. This would therefore require the 3D MOT to be carefully placed off the beam-axis and for the vacuum chamber to incorporate potentially non-trivial geometries. This is a known complication when trying to load directly on the intercombination transition of Yb systems~\cite{Wodey_2021}. We circumvent this problem by instead separating the two MOT chambers along the vertical axis, exploiting the acceleration due to gravity to help the atoms fall towards the 3D-MOT region~\cite{Bennetts_2017}. This has the additional benefit of reducing the required power of the push beam, helping to protect the intercombination-transition MOT from the more powerful dipole-transition light.

The loading of the 3D MOT will also be enhanced by including two additional stages of cooling on the 326~nm transition: firstly in the transverse direction in optical molasses; and then using a pair of angled and crossed beams to slow in the longitudinal (vertical) direction.
The combination of the relatively fast transverse velocity due to the high Doppler temperature of the $^1$S$_0$-$^1$P$_1$ transition with the slow longitudinal velocity results in a divergent atomic beam ($\sim$~100~mrad). The transverse cooling is therefore required for collimation of the slow atomic beam coming from the 2D MOT, without which only a small fraction of the atoms would be capturable in the 3D MOT. The additional longitudinal slowing is less critical, but allows for a greater fraction of atoms to be captured by reducing the vertical component of the velocity gained during free fall. These beams are so-called crossed (or angled) slowing beams, a geometry which avoids interference with the 3D MOT itself and as has been demonstrated effectively in e.g. Dy, Er and Yb systems~\cite{Lunden_2020,Seo_2020,Plotkin-Swing_2020}.

\begin{figure*}[t]
\centering\includegraphics[width=0.93\textwidth]{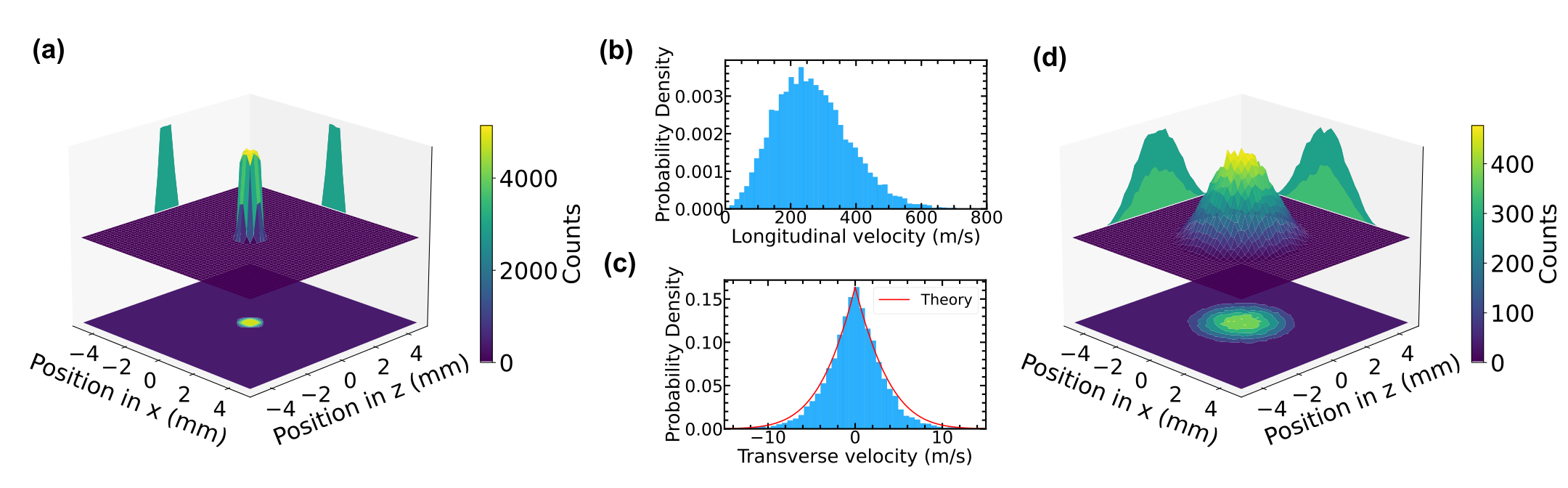}
\caption[Simulated output of the cadmium oven]{Velocity and position distributions (n=10$^4$) simulated at the output of our oven at T=100~$^\circ$C. (a) Positional distribution in the transverse beam plane ($x$-$z$, see Fig.~\ref{fig:2dMOT}); (b) longitudinal velocity distribution; (c) transverse velocity distribution, compared to the theoretical distribution. (d) Positional distribution in the transverse beam plane at an axial distance of 75 mm from the oven -- the approximate position of the 2D MOT where most of the atoms are within a $\sim$2~mm radius.}
\label{fig:ovenBeam}
\end{figure*}

\section{Numerical Simulation of Atomic Trajectories}\label{sec:simulation}
We numerically simulate the atomic trajectories of our system using pseudorandom sampling and the Monte Carlo method, implementing this process on \textsc{Python}. This technique has been successfully applied previously to simulate a broad variety of MOTs and MOT-based atomic beam sources, including standard 3D and 2D implementations~\cite{Wohlleben_2001,Chaudhuri_2006,Barbiero_2020}, to more unconventional configurations such as pyramidal~\cite{Kohel_2003} and Rydberg-dressed systems~\cite{Bounds_2018}. 

The atomic trajectories are determined for a pseuedorandomly drawn initial position and velocity and by stepping the time sequentially and using the calculated acceleration due to radiation pressure to update the atom's position and velocity for the next time step. Although it is possible to perform a quantum full simulation of MOT dynamics~\cite{Hanley_2018}, we instead chose temporal step sizes $\tau$ such that $\tau > 1/\Gamma$ so that the atom-light interaction can be treated in a semi-classical manner. As different regimes of the simulation are dominated by different transitions with highly different linewidths, we alter this time step accordingly and also in a trade-off between accuracy and computational time, but $\tau$ is typically $\sim$50~$\mu$s. The total end time for the simulation is made longer than generally required ($\sim$500~ms), with the simulation of each atom instead stopped when it fulfils certain criteria, such as leaving a certain spatial range or becoming trapped in the MOT.

The starting point of our simulation is an effusive atomic beam. Collimated sources of Cd have previously been demonstrated, continuously with a capillary-based oven system with a divergence $\sim$40~mrad~\cite{Tinsley_2021}, and in a pulsed manner using laser ablation~\cite{Ohayon_2022}. Here we model an oven with a simple single 32-mm long, 1-mm diameter capillary, for which the Knudsen number $K_N\gg$1 at 100~$^\circ$C, taking a Van der Waals radius of 158~pm for Cd~\cite{Bondi_1964}, meaning intra-atomic collisions can be ignored. Although the longitudinal and transverse velocity distributions from a capillary-based oven are known~\cite{Schioppo_2012,Gao_2014}, it is not possible to sample from them independently, due to the permissible range of transverse velocities for successfully exiting the capillaries depending upon the longitudinal velocity. We instead use the Monte-Carlo method to generate three velocity components using the Maxwell-Boltzmann distribution and geometrically determine whether these atoms will exit our oven design. This simulation is performed until the desired number of atoms have successfully exited the capillary, which is typically 10$^4$. The generated transverse velocity distributions of this simulation match the theoretical distribution well~\cite{Gao_2014,Greenland_1985}, as shown in Fig.~\ref{fig:ovenBeam}. 

All the laser beams are modeled as perfect Gaussian beams, with a sharp truncation introduced by the diameter of the vacuum viewport they will be shone through. We determine the local light-induced acceleration by these beams at each position and velocity by considering a model that includes the vector of the local magnetic field, not just the field magnitude~\cite{Vangeleyn_2009}. In this model, the polarisation of the light is decomposed into its different $\sigma^-$, $\pi$ and $\sigma^+$ components, following the quantisation axis provided by the local magnetic field direction. 
This allows for a more accurate determination of the scattering force at arbitrary 3D fields and positions within our simulation, for example when the atom is not along the beam axes. Finally, when simulating the $^1$S$_0$-$^3$P$_1$ transition, we typically assume that the laser beam is frequency modulated. This technique is often used to enhance the trapping potential of MOTs on intercombination transitions~\cite{Kuwamoto_1999}. We model this by assuming the total power of the laser beam to be evenly distributed between the $j$ frequency modes.

\begin{figure*}[!t]
\centering\includegraphics[width=\textwidth]{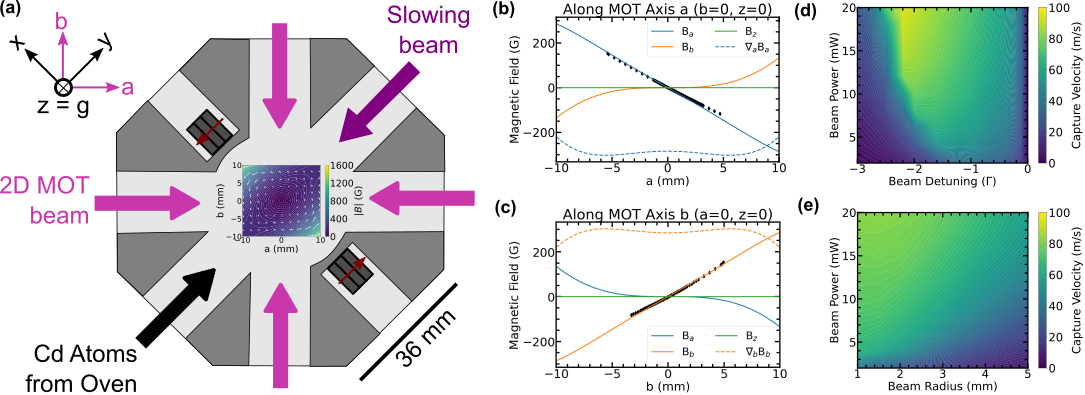}
\caption[Design of the compact cadmium 2D MOT]{The Cd atom source. (a) To-scale drawing of the 2D-MOT chamber design and the field generated by the permanent magnets. (b) and (c) Calculated magnetic field components (solid lines) along the MOT beam axes. For the relevant component along the beam axis, the calculated field gradient (dashed line, units G/cm with the same scale) and the measured magnetic field (black diamonds) are also shown. Contour plots of the simulated capture velocity of the 2D MOT as a function of (d) beam detuning and beam power ($w= 2$~mm) and as a function of (e) beam power and radius ($\Delta$=$-$1.5$\Gamma$) are also shown. The atoms slowed in the 2D MOT are accelerated vertically downwards by a weak push beam (not shown).}
\label{fig:2dMOT}
\end{figure*}

Following this formalism~\cite{Vangeleyn_2009} and adjusting for the possibility for multiple frequency modes, we can write the acceleration due to the $i^{\text{th}}$ laser beam, propagating in the direction of a unit vector $\hat{\textbf{k}_i}$ with intensity $I$ and operating on a transition with natural linewidth $\Gamma$ and vacuum wavelength $\lambda$:
\begin{multline}
\label{eq:lightAcc}
    a_i = \frac{h\pi\Gamma}{m\lambda}s_j\hat{\textbf{k}_i}\cdot\\
    \sum^{j} \sum_{n=-1,0,1}\frac{\eta_n}{1+s_{\text{tot}}+4\left(\Delta_\Gamma-k_\Gamma\right)\hat{\textbf{k}_i}\cdot\textbf{v}-\Delta g_{F}\mu_\Gamma n\left|\textbf{B}\right|},
\end{multline}
where $s_j=I/jI_\text{sat}$ is the saturation parameter of a single frequency mode and $s_{\text{tot}}=\sum_is_j$ is the combined saturation parameter from all beams~\cite{Vangeleyn_2009,Ovchinnikov_2005}, $\Delta_\Gamma$ is the detuning of the mode in units of linewidth, $k_\Gamma=1/\left(\lambda\Gamma\right)$, $\mu_\Gamma=\mu_B/\left(2\pi\Gamma\right)$ with $\mu_B$ the Bohr magneton, and $m$ is the atomic mass, $\textbf{v}$ the atomic velocity and $\hat{\textbf{B}}$ is unit vector of the magnetic field at the position in question. We concentrate on the bosonic isotopes of Cd for which $\Delta g_{F}=g'_Fm_F'-g_Fm_F$ is 1 and $1/6$ for the 229~nm and 326~nm MOT transitions, respectively. The $\sigma^-$, $\pi$ and $\sigma^+$ components are accounted for by the summation over $n$ and with $n$ = -1, 0, 1, respectively. The parameter $\eta_n$ which is given by $\eta_0=\left(1-\left(\hat{\textbf{k}_i}\cdot\hat{\textbf{B}}\right)^2\right)/2$ and $\eta_{\pm1}=\left(1\mp \alpha\hat{\textbf{k}_i}\cdot\hat{\textbf{B}}\right)^2/4$, where $\alpha=\pm$1 is the handedness of the circularly polarised light relative to the propagation direction~\cite{Vangeleyn_2009}. We also extend this formalism to account for linear polarisations, as well as circular. In this case we instead use $\eta_0=\left(\hat{\textbf{E}_i}\cdot\hat{\textbf{B}}\right)^2$ and $\eta_{\pm1}=\left(1-\left(\hat{\textbf{E}_i}\cdot\hat{\textbf{B}}\right)^2\right)$/2, where $\hat{\textbf{E}_i}$ is the unit linear polarisation vector of the beam.

The details of the various magnetic field calculations are given in the relevant sections below, but in all cases we first calculate a field on a spatial grid across the full experimental region. These calculations are then linearly interpolated and saved for computational efficiency, allowing for the determination of the total magnetic field from all sources at any spatial points within the simulation region. Combined with Eq.~\ref{eq:lightAcc}, this means that the we can simulate the light force at all points. Earth's magnetic field is assumed to be consistent and cancelled and therefore not considered.

In addition to determining the acceleration due to each beam, we also use Eq.~\ref{eq:lightAcc} to estimate the total scattering rate $R$. This can then be used to model the heating effect of spontaneous emission via the addition of a random momentum kick $\hbar\left|\textbf{k}\right|\sqrt{R\tau}\hat{\textbf{x}}$, where $\hat{\textbf{x}}$ is a unit vector chosen pseudorandomly from an isotropic distribution~\cite{Kohel_2003,Barbiero_2020}. In this way, the temperature of the atoms is limited to the Doppler temperature~\cite{Letokhov_1977} instead of continuing to decrease towards zero, which is important for correctly understanding the behaviour of the $^1$S$_0$-$^1$P$_1$ transition stages (Sec.~\ref{sec:2dmot}) where the 2.2~mK Doppler temperature leads to non-negligible residual velocities.

\section{The 2D-MOT, Zeeman Slower and Push Beam at 229~nm}\label{sec:2dmot}

The first-stage of cooling and trapping of our design is to load atoms from the effusive beam into a 2D MOT. We have designed a compact chamber with cut-out regions that allow for permanent magnets to be placed at a minimum distance of 22~mm from the MOT centre, but to remain external to the vacuum chamber for experimental ease (Fig.~\ref{fig:2dMOT}). With two stacks of three permanent bar magnets each (neodymium, 25$\times$10$\times$3~mm$^3$, M$\sim$9$\times$10$^5$~A/m), we can generate magnetic field gradients of $\sim$250~G/cm which are approximately uniform across the MOT region. A set of three magnets is the maximum possible with our chamber design and it produces a higher capture velocity than a set of one or two. Although analytic solutions for the field produced by bar magnets exist~\cite{Cheiney_2011}, we find minimal deviations when modelling the magnets as point-source dipoles. Fig.~\ref{fig:2dMOT} shows both the calculated and measured magnetic field profiles, as determined with a Hall probe, which are in good agreement.

\begin{figure}[t]
\centering\includegraphics[width=0.5\textwidth]{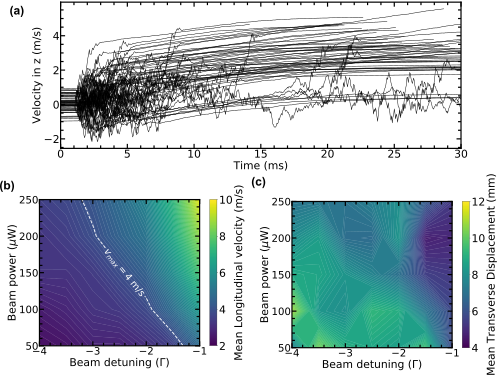}
\caption[Low-intensity push for generating a slow beam of atoms]{The low-intensity push for generating a slow beam of atoms. (a) Sample trajectories of atoms interacting with the 2D MOT and push beam, most interactions happen within a 5~ms time frame. (b) Mean longitudinal velocity of the atoms exiting the 2D MOT (50~mm distance) as a function of push beam power and detuning ($w$=3~mm); (c) the mean transverse displacement of the atoms from the beam axis for the same variables. Only around 150 $\mu$W of power is required to produce atoms at the target velocity of 4~m/s.}
\label{fig:pushBeam}
\end{figure}

We can use this field to estimate the capture velocity, to the nearest m/s, of the 2D MOT using the simulation method outlined in the previous section. In these simulations we use atoms without transverse velocity and remove the random heating arising from scattering, effectively therefore only considering cooling and trapping in 1D. We perform this simulations for a range of beam powers, beam radii, and frequency detunings, showing that capture velocities approaching 100~m/s are achievable for this configuration (Fig.~\ref{fig:2dMOT}). We find that for a reasonable 229~nm power of just 10~mW per beam and with a beam radius of 2~mm and a detuning of $-$1.5~$\Gamma$, this configuration can achieve capture velocities of $\sim$70~m/s. This beam radius is chosen to match the atomic beam size at the 2D MOT position (Fig.~\ref{fig:ovenBeam}~(d)). The high capture velocity is a positive feature of the 229-nm transition, arising from the large accelerations achievable due to the low wavelength and high linewidth (see Table~\ref{tab:atomProperties}). Although increasing the power can improve the performance (Fig.~\ref{fig:2dMOT}), we limit the power to 10~mW to be comfortable for long-term production of current laser technology~\cite{Tinsley_2021} and to protect the vacuum viewports. Likewise we use a detuning of $-$1.5~$\Gamma$ to be within a region which is immune to frequency and intensity fluctuations, as the capture velocity drops dramatically with increasing detuning or declining intensity following the optimum value. The resulting $\sim$70~m/s capture velocity will allow for appreciable atom numbers to be loaded into the 2D MOT directly from an effusive beam or vapour, as studied later in Section~\ref{sec:vacuum}.

Following trapping in the 2D MOT, we can consider the case of a low-intensity push beam orthogonal to the 2D-MOT plane, which serves the purpose to plug the atoms in one direction and to accelerate them in the other, generating a slow beam of atoms. This beam is low-intensity to reduce the 229-nm power requirements, to maintain a small velocity of the output atoms, and to not interfere with the 326-nm MOT which is placed directly vertically below (see Section~\ref{sec:3dmot}). By incorporating the push beam, we can simulate the interaction time with the 229~nm light in the 2D MOT and push beam, finding it to be just a few ms (Fig.~\ref{fig:pushBeam}~(a)). For comparison, a steady-state 3D MOT on this transition with similar beam parameters is loaded for at least 200~ms~\cite{Yamaguchi_2019}. The expected losses due to photoionisation for our system are therefore around only 2\%, given a calculated value of $\Gamma_{\text{ion}}$=4~Hz, and such losses can be considered negligible.
\begin{figure}[t]
\centering\includegraphics[width=0.5\textwidth]{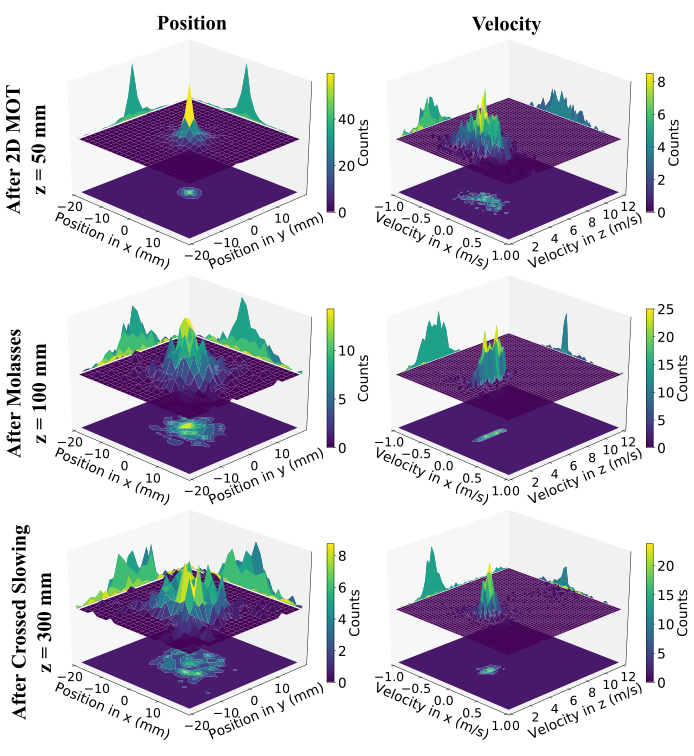}
\caption[Evolution of the slow cadmium beam]{Simulated evolution of the position (upper row) and velocity (lower row) distributions of our slow atomic beam as it propagates in the vertical direction $z$. Columns from left to right show the distributions at a distance of 50~mm below the 2D MOT; 100~mm below the 2D MOT and after transverse cooling at 326~nm; and 300~mm below the 2D MOT and after the crossed slowing beams. A positive value of $v_z$ is in the direction of gravity, the effect of which is included.}
\label{fig:slowBeam}
\end{figure}

\begin{figure*}[t]
\centering\includegraphics[width=0.96\textwidth]{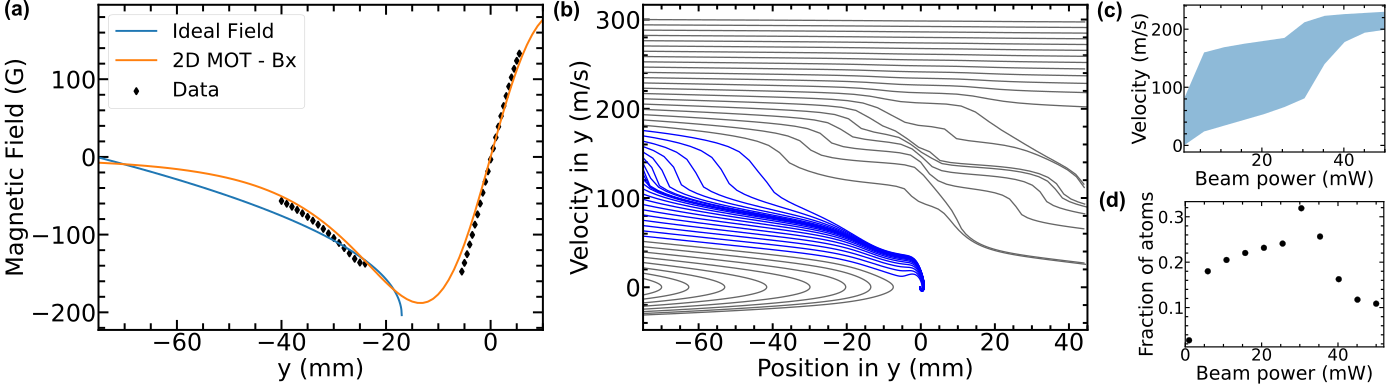}
\caption[Zeeman slower for cadmium based on permanent magnets]{(a) Blue line shows the ideal field of the Zeeman slower for slowing atoms from 135~m/s to 75~m/s over 58~mm with $\Delta$ = -6.5~$\Gamma$. The orange line shows the transverse field from the 2D MOT magnets and the black diamonds the data measured with a Hall probe. (b) The longitudinal velocity of atoms coming from the oven to the 2D MOT when the Zeeman slower beam power is set to 20~mW. Blue traces are those captured in the 2D MOT. (c) The shaded region shows the oven output velocities captured in the 2D MOT when adding a Zeeman slowing beam. (d) The corresponding fraction of the oven output at 100 $^\circ$C. This value peaks for 30~mW of beam power at around 30\% of atoms, up from 3\% for the case without any slower.}
\label{fig:zeemanSlower}
\end{figure*}

To investigate the output atomic beam we simulate atoms from our atomic oven which are then slowed and trapped in the 2D-MOT and exit along the axis of the push beam for a range of experimental powers and frequency detunings, with the beam radius fixed at 3~mm to be larger than the size of the 2D MOT. In Fig.~\ref{fig:pushBeam} we consider the output velocity and positional spread of such atoms at a vertical distance of $z=50$~mm from the 2D-MOT centre, with a target axial velocity of around 4~m/s. At this distance, the force from the push beam has become small and the velocity distribution is largely fixed (Fig.~\ref{fig:pushBeam}~(a)). The longitudinal velocity of the output atomic beam is controlled by the intensity of the push beam and its detuning, as shown in Fig.~\ref{fig:pushBeam}~(b). The advantage of increasing the frequency detuning is that the sensitivity to power fluctuations is reduced, for example, decreasing the sensitivity by a factor 2 when increasing the detuning from $-$2$\Gamma$ to $-$3$\Gamma$. Even at $\Delta$=$-$3$\Gamma$, however, only $\sim$150~$\mu$W is required for $v_z$=4~m/s. 
For the specific case of a push beam power of 170~$\mu$W and $\Delta$=$-$3$\Gamma$, the simulated distribution is shown in Fig.~\ref{fig:slowBeam}, giving an axial velocity centred around 4~m/s with a transverse spread of $\pm$0.5~m/s, compatible with the Doppler temperature.
With such a velocity distribution, the mean transverse displacement from the vertical axis also remains modest over short distances ($<$10~mm at 50~mm displacement), though subsequent transverse cooling on the narrow 326-nm intercombination transition is mandatory, as discussed later (Section~\ref{sec:3dmot}).

We finally consider the possibility of enhancing the atom number in the 2D MOT by use of a Zeeman slowing beam. 
Generating these fields over the short distances requested ($\sim$60~mm) is challenging, even with permanent magnets. For example, while permanent magnets in Halbach arrays have been used and studied in detail for Zeeman slowers with e.g. Rb~\cite{Cheiney_2011,Ali_2017} and Yb~\cite{Wodey_2021}, the generated fields had gradients of 3~G/cm~\cite{Cheiney_2011}, 12~G/cm~\cite{Ali_2017} and 20~G/cm~\cite{Wodey_2021}, an order of magnitude lower than what is requested in this case ($\sim$100~G/cm). Such gradients are difficult to design, especially without affecting the magnetic field in the 2D MOT region.

However, the negative gradient slope of the 2D MOT field makes a reasonable approximation of a transverse field Zeeman slower, as shown in Fig.~\ref{fig:zeemanSlower}~(a). 
This field requires linear polarisation orthogonal to the magnet field direction~\cite{Ovchinnikov_2007}, which is therefore equally decomposed into $\sigma^+$ and $\sigma^-$ components. Only half the input power is therefore available to drive the $\sigma^+$ needed for our decreasing field configuration, effectively doubling the power requirements compared to a longitudinal field Zeeman slower~\cite{Hill_2014}, which runs counter to the design idea of minimizing the required 229-nm power (see Section~\ref{sec:overview}).

Nevertheless, the performance of a Zeeman slower using this field is shown in Fig.~\ref{fig:zeemanSlower} as a function of the slowing beam power. The beam waist is 2~mm (focused at the oven output) and the detuning $\Delta= -6.5$~$\Gamma$ to match the ideal field as closely as possible. The range of oven output velocities captured by the 2D MOT is shown, as atoms that are too slow can be pushed backwards by the Zeeman slower beam, especially at higher powers (Fig.~\ref{fig:zeemanSlower}~(c)). This range can be approximately converted into a normalized atom number by integrating the longitudinal velocity distribution of the oven output (see Fig.~\ref{fig:ovenBeam}) within the capture velocity range. As can be seen in Fig.~\ref{fig:zeemanSlower}~(d), at $\sim$30~mW of power the fraction of atoms it is in principle possible to load into the 2D MOT is increased to $\sim$30\%, an order or magnitude improvement from the case without the Zeeman slower. Due to this large required beam intensity, the Zeeman slower is in general ignored in the subsequent analysis and discussion.

\section{Direct Loading of a 3D-MOT at 326~nm}\label{sec:3dmot}

Whilst the broad dipole-allowed $^1$S$_0$-$^1$P$_1$ transition allows for efficient slowing of the fast atoms from the longitudinal beam, the Doppler temperature of 2.2~mK is impractical for either direct performance of atom interferometry or for further cooling in an optical dipole trap (see Section~\ref{sec:odt}). Further cooling with the $^1$S$_0$-$^3$P$_1$ transition is therefore required, as discussed in Section~\ref{sec:coldCd}, and we propose to directly capture the slow atomic beam exiting the 2D MOT into a MOT based on this transition.

To understand the feasibility of this approach, we first numerically determine the capture velocity of a Cd MOT at 326~nm, for a broad range of experimentally realistic parameters. In simulating this process, we consider copper-wire coils wound around a standard DN100CF flange (see Section~\ref{sec:vacuum} for vacuum system details), and calculate the generated magnetic field analytically along the axial direction and numerically otherwise, and perform a linear interpolation between these points. We assume usable laser powers of up to 100~mW per beam in a retro-reflected configuration, based on our recently developed system~\cite{Manzoor_2022}.
\begin{figure}[t]
\centering\includegraphics[width=0.5\textwidth]{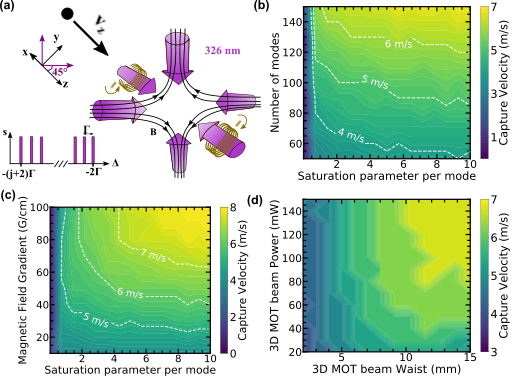}
\caption[Capture velocity of the 3D MOT at 326~nm]{(a) Schematic of the simulation of the 3D MOT at 326~nm. Atoms downwards along the vertical $z$ axis and interact with beams propagating along the $x$ axis and at 45$^\circ$ to the $y$ and $z$ axes ($x=0$ plane). The magnetic field coils produce an axial field along the $x$ axis. Frequency modulation of the $j$ modes is simulated assuming the first mode at $\Delta$=$-2\Gamma$ with subsequent modes further detuned by $\Gamma$ and all modes having the same saturation parameter $s$. The capture velocity is determined as a function of (b) saturation parameter per frequency mode and magnetic field gradient ($w$=5~mm), (c) the saturation parameter per frequency mode and the number of modes ($\nabla B$=30~G/cm, $w$=5~mm), and (d) the beam power and beam waist ($\nabla B$=30~G/cm, modes = 100). For a broad range of feasible experimental parameters, a capture velocity of around 5~m/s is achievable.}
\label{fig:3dMOTCaptureV}
\end{figure}

We determine the the capture velocity for a broad range of parameters for atoms travelling along the vertical $z$ direction and with the MOT beams propagating through the coils along the $x$ axis and the other beams at 45$^\circ$ to the $y$ and $z$ axes in the $x=0$ plane. (Fig.~\ref{fig:3dMOTCaptureV}~(a)). Figures~\ref{fig:3dMOTCaptureV}~(b) and~(c) show that the capture velocity can be $>$5~m/s for a large range of feasible parameters in terms of magnetic field gradients and frequency modulation given the available power. In the case of a magnetic field gradient $\nabla B$=30~G/cm, as used previously for this MOT~\cite{Yamaguchi_2019} and 100 frequency modes evenly separated by $\Gamma/2\pi$ (6.6~MHz amplitude), we find that beams with radii $\geq$5~mm are required to capture atoms at $v_z$=5~m/s (Fig.~\ref{fig:3dMOTCaptureV}~(d)).

\begin{figure}[t]
\centering\includegraphics[width=0.5\textwidth]{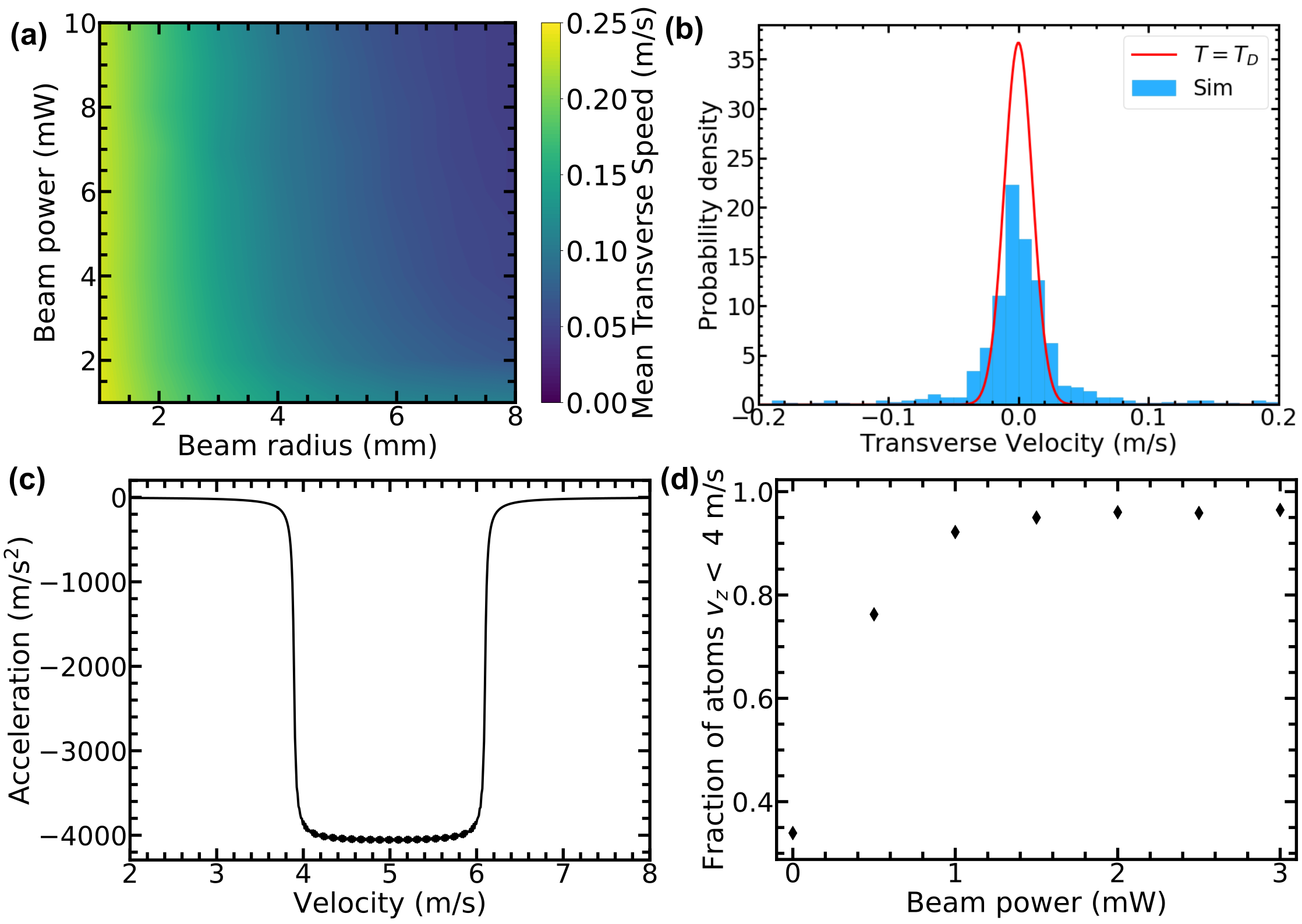}
\caption[Transverse and longitudinal cooling at 326~nm of the slow atomic beam]{Transverse and longitudinal cooling at 326~nm of the slow atomic beam. (a) Mean transverse velocity of the slow atomic beam following the 326-nm optical molasses for a range of beam radii and powers (100 frequency modes). (b) Sample transverse velocity distribution after the molasses stage ($w$=5~mm, $P$=10~mW) showing that the output approaches the Doppler limit. (c) Longitudinal deceleration on the atoms due to the crossed slowing beams angled at 16$^\circ$, targeting slowing down to 4~m/s. (d) Fraction of the atoms having a longitudinal velocity less than 4~m/s following the crossed slowing.}
\label{fig:molasses}
\end{figure}

\begin{figure*}[t]
\centering\includegraphics[width=0.97\textwidth]{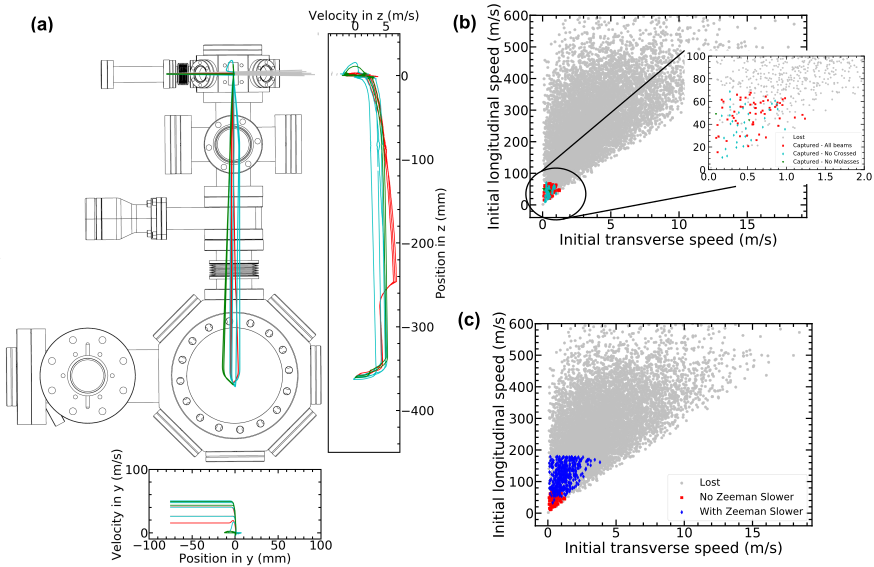}
\caption[Vacuum chamber design and simulation of cadmium trajectories]{(a) To-scale drawing of the vacuum chamber design overlaid with a sample of simulated atomic trajectories. Some features, such as vacuum pumps, have been removed for clarity. The colours of the trajectories represent different outcomes: grey lines are always lost, green lines are captured with just the 2D and 3D MOTs active, cyan lines are captured when the transverse cooling is also activated, and red lines when the crossed slowing beams are active. See text for details. (b) Atoms captured for different cooling configurations based on velocity output from the oven, with the same colour scheme as (a). (c) Atoms captured with (blue diamonds) and without the (red squares) the Zeeman slower stage being active.}
\label{fig:vacuum}
\end{figure*}
 
While the first-stage of cooling produces a beam of atoms sufficiently slow to be captured by this MOT (see Section~\ref{sec:2dmot}), a problem arises in the that the transfer distance from the 2D to 3D MOT will be $\sim$35~cm, based on standard vacuum components and other considerations such as differential pumping. Due to the slow longitudinal velocity and high Doppler temperature of the 229~nm transition, the atoms will diverge significantly over this distance (see Fig.~\ref{fig:slowBeam}). We therefore simulate transverse cooling 85~mm below the 2D MOT in 2D optical molasses with the $^1$S$_0$-$^3$P$_1$ transition, which effectively collimates the slow atomic beam. 
The slow atomic beam after the molasses is shown in Fig.~\ref{fig:slowBeam} for the case of 10~mW per 5~mm radius beam with 100 frequency modes. The transverse velocity spread has been reduced to $<\pm$0.05~m/s, an order of magnitude reduction. The lack of cooling in the longitudinal direction does cause some heating, though this is not seen to be significant. 

Furthermore, as shown in Fig.~\ref{fig:molasses}~(a), this cooling is effective for a broad range of beam powers and radii, requiring a beam radius $>$4~mm and powers $>$2~mW to approach optimal transverse cooling. The transverse velocity distribution approaches the Doppler limit (Fig.~\ref{fig:molasses}~(b)) and substantial transverse cooling can be achieved for even low powers. Moreover, we find that this cooling can occur relatively rapidly over just 2~ms, with reasonable power levels (10~mW per beam, radius 5~mm) and frequency modulations (100~modes). This cooling is also stronger than the contribution of the 229~nm scattered from the 2D MOT. To estimate the effect of this on the atoms, we consider a worst-case scenario of all the input light being scattered resonantly and isotropically from the MOT centre. At the molasses position, the force from this scattered light is two orders of magnitude lower than the force from a single beam of the molasses and it is therefore taken to have a negligible effect.

\begin{figure*}[t]
\centering\includegraphics[width=\textwidth]{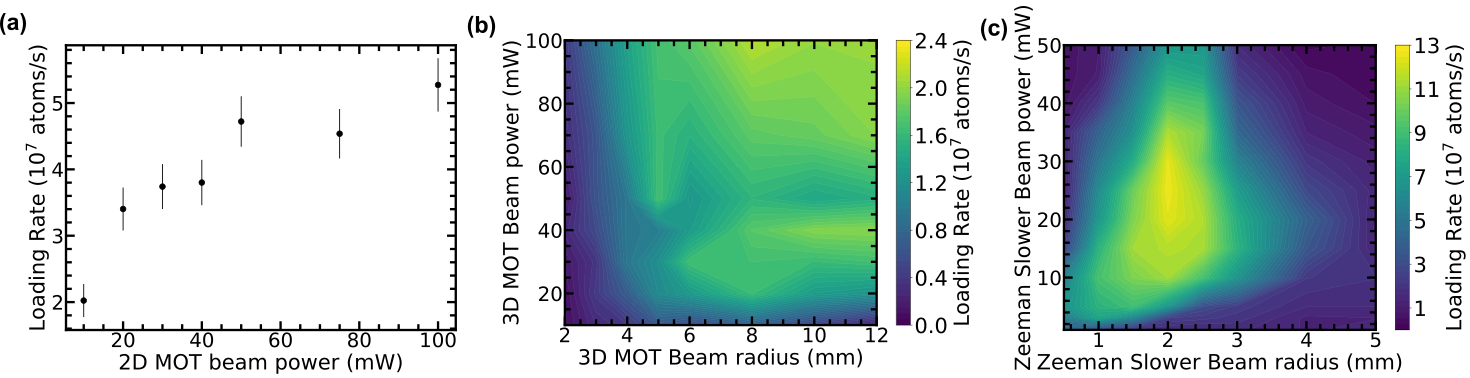}
\caption[Simulated loading rate of $^{114}$Cd into the 3D MOT]{Simulated loading rate of $^{114}$Cd into the 3D MOT. (a) Loading rate as a function of 2D MOT beam power. (b) Loading rate as a function of 3D MOT beam power and radius (100 frequency modes). (c) Loading rate as a function of Zeeman slower beam power and radius. All simulations with 10$^4$ atoms and non-variable parameters as shown in Table~\ref{tab:beamProperties}, except the Zeeman slower which is only used in (c).}
\label{fig:loading}
\end{figure*}

The capture efficiency of the MOT can be further enhanced by using vertical slowing on this output of the 2D molasses. We consider a pair of crossed slowing beams aligned with a full angle of 16$^\circ$ and again on the 326~nm transition. Assuming 100 frequency modes, the acceptance velocity range of the force is $\sim$2~m/s (Fig.~\ref{fig:molasses}~(c)), meaning the detuning $\Delta$ must be carefully selected in order to slow the desired atoms, also accounting for the non-negligible Zeeman shifts arising from the MOT coils (see Eq.~\ref{eq:lightAcc}). In our case this corresponds to a detuning of $\Delta$=$-$650~$\Gamma$ (43~MHz) to slow the atoms to a minimum velocity of 4~m/s. With just a few mW of power, the crossed slowing of atoms increases the fraction of atoms with a longitudinal velocity $v_z<$4~m/s from 30\% to 95\% (Fig.~\ref{fig:molasses}~(d)). As shown in Fig.~\ref{fig:slowBeam}, this slowing process produces only minimal heating in the transverse direction and the slow atomic beam from the 2D MOT has therefore been both collimated for efficient transport and further cooled to below the 3D MOT capture velocity.

\section{Full Vacuum Apparatus \& MOT Loading Rates}\label{sec:vacuum}

We utilise the results of the preceding sections to design a vacuum system capable of generating a cold beam of Cd which can be loaded into a MOT on the 326-nm $^1$S$_0$-$^3$P$_1$. A to-scale diagram of the design is shown in Fig.~\ref{fig:vacuum}. The 2D and 3D MOT regions are each pumped with an ion pump and non-evaporable getter and are separated by a gate valve to allow for the replacement of viewports in the case of UV-induced damaged, without having to open the whole system. This geometry also provides the finalised distances for the different cooling stages, namely the molasses cooling 85~mm below the 2D MOT and the crossed slowing 100~mm above the 3D MOT, which is itself 350~mm below the 2D MOT. With this finalised design, we are able to simulate the atomic trajectories throughout the whole system at a range of oven temperatures and estimate the atomic loading rate into the 3D MOT, a more useful experimental parameter than e.g. the capture velocity used above. 

Information on all the simulated beam powers, radii, frequency detuning etc. is given in Table~\ref{tab:beamProperties}. We first attempt to quantify the effect of the different stages of our design, by running the simulation at $T$=100~$^\circ$C (partial pressure of 2.5$\times$10$^{-7}$~mbar, flow rate of 1.1$\times$10$^{10}$~atoms/s) with and without the Zeeman slower and the crossed (vertical) and transverse cooling at 326~nm. We find that the introduction of the transverse slower increases the capture efficiency by more than an order of magnitude and with the introduction of vertical slowing it increases by another factor $\sim$3 (Table~\ref{tab:motLoading}), for an approximate total factor of 40 in the efficiency. As Fig.~\ref{fig:vacuum}(b) shows, with these two cooling stages active, nearly all of the atoms below the capture velocity of the 2D MOT are captured by the 3D MOT, showing that the transfer between the MOTs is highly efficient. Adding the Zeeman slower leads to a substantial increase in the capture efficiency, by capturing a higher initial longitudinal velocity class in the 2D MOT, as shown in Fig.~\ref{fig:vacuum}(c).

As shown in Table~\ref{tab:motLoading}, we estimate expected loading rates into our 3D MOT and find values $\sim$10$^7$~atoms/s for an oven temperature of 100~$^\circ$C, without the Zeeman slower. When adding the Zeeman slower beam, this increases by approximately a factor 5 to 10$^8$~atoms/s, though at the cost of a substantial increase in the required 229~nm power. The loading rate is determined by calculating the expected flow rate of atoms at the simulated oven temperature and capillary design and then multiplying this by simulated capture efficiency. We further scale the loading rate based on the fractional natural abundance of the $^{114}$Cd isotope of 0.29~\cite{Berglund_2011}. Although we have designed the system to work with a minimal amount of 229~nm light, we note that the system is scalable should problems such as stable power production and vacuum viewport damage be solved (see Section~\ref{sec:coldCd}). In addition to allowing for a Zeeman slower, this would allow for an increase in the 2D MOT beam powers, resulting in an approximate threefold increase in loading rate (Fig.~\ref{fig:loading}).

The parameters presented in Table~\ref{tab:beamProperties} can also be varied to look for the optimum loading rate, especially the effect of the 3D MOT beams and the Zeeman slower. Figure~\ref{fig:loading} shows that the loading rate is robust for a broad range of 3D MOT beam powers and beam radii (100 frequency modes), provided the power is $>$10~mW and the beam radius $>$3~mm. The Zeeman slower beam shows a maximum loading rate which differs slightly from what would na\"ively be expected from looking only at the velocity class addressed by the slowing beam (cf. Fig.~\ref{fig:zeemanSlower}~(d) and Fig.~\ref{fig:loading}~(c)). This is due to the small force imbalance the Zeeman slower introduces to the 2D MOT, which can deflect the slow atomic beam off axis, especially when the Zeeman slower beam radius exceeds the 2D MOT beam radius. Substantial increases in loading rates are available (factor 5), even for powers down to around 10~mW with a focused beam ($w<$2~mm).

It should be noted that these loading rates represent the upper bound for the loading rate as our single-atom simulation does not consider losses such as collisions with background gases or other cold Cd atoms, as well as photoionisation losses. 
Nevertheless, they suggest that significant numbers of atoms can be quickly loaded into an intercombination transition MOT without the need for significant powers or interaction times with the problematic 229~nm $^1$S$_0$-$^1$P$_1$ transition. The loading rates presented can also be enhanced by increasing the oven temperature above the modest 100~$^\circ$C used here, and by an approximate factor 3 by using enriched cadmium sources, as have been employed previously elsewhere~\cite{Yamaguchi_2019,Hofsass_2023}.

\begin{table}[t]
 \caption{Properties of the laser beams in the final simulation of the system, including optical power (per beam) $P$, beam waist $w$, detuning $\Delta$, number of frequency modes and the saturation parameter per mode $s$.} 
 \label{tab:beamProperties}
 \begin{center} 
 \begin{tabular}{l c c c c c}
    \hline
    \hline
    Beam   (number)       & $P$ (mW)  &  $w$ (mm)       & $\Delta$ ($\Gamma$)     & Modes & s \\
    \hline
    \multicolumn{6}{c}{229 nm -- $^1$S$_0$-$^1$P$_1$}\\
    \hline
       2D MOT ($\times$4)&   10         &       2          & -1.5                & 1 & 0.16\\
       Push beam ($\times$1)&   0.17        &     3             & -3               & 1 & 10$^{-3}$\\  
       Zeeman Slower ($\times$1) &   25         & 2             & -6.5               & 1 & 0.4\\
    \hline 
    \multicolumn{6}{c}{326 nm -- $^1$S$_0$-$^3$P$_1$ }\\
    \hline
       3D MOT ($\times$6)&  50          &        5         & -2            & 100 & 5.1\\
       Transverse Cooling ($\times$4)&   10         &        5          & -1   & 100 & 1.0\\
       Crossed Slowing ($\times$2)&   2.5         &  3                & -650   & 100 & 0.71\\
    \hline     
    \hline
 \end{tabular}
 \end{center}
\end{table}

\section{Trapping, Transfer \& Launching at 1064~nm}\label{sec:odt}

Although the micro-Kelvin temperatures achieved in the 326-nm MOT~\cite{Yamaguchi_2019} are sufficient for many applications, producing quantum degenerate sources requires further cooling towards the nK level. This is typically achieved by performing evaporative cooling in an optical dipole trap~\cite{Adams_1995,Barrett_2001}, neither of which techniques have been demonstrated with Cd yet. In this section we consider the feasibility and prospects of this approach and also the discuss the transfer and launching of Cd atoms using related techniques, with particular reference to a dual-species interferometer with Sr~\cite{Tinsley_2022}.

\begin{figure*}[t]
\centering\includegraphics[width=\textwidth]{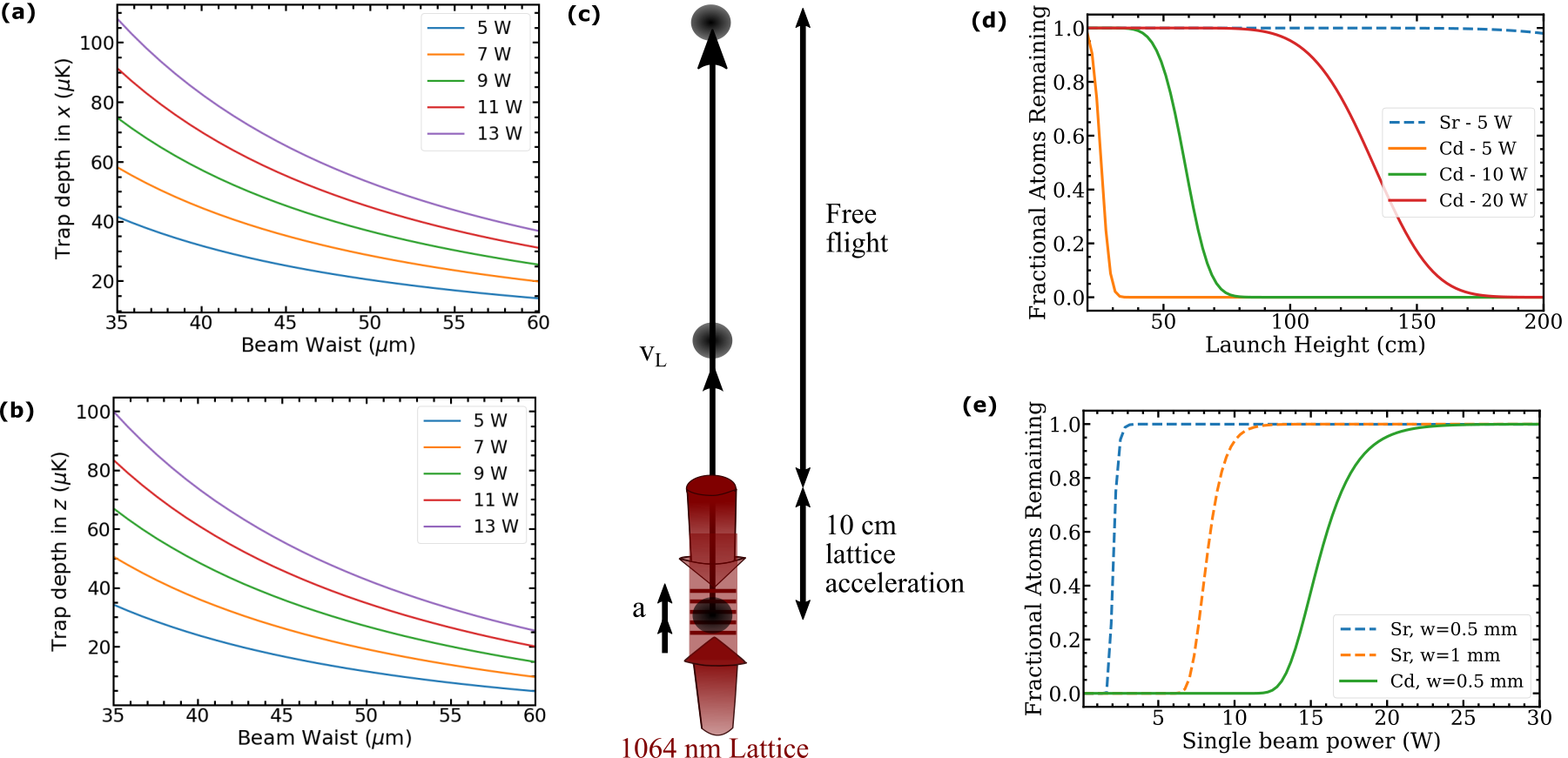}
\caption[Prospects for dipole trapping and lattice launching at 1064~nm]{Prospects for dipole trapping and lattice launching at 1064~nm. Calculated depth of the optical dipole trap in the (a) horizontal and (b) vertical directions as a function of single-beam power and waist. Depths significantly larger than the 3D MOT Doppler temperature are available. (c) Proposed launch scheme in which atoms are accelerated in a moving lattice over a distance of 10~cm and released with velocity $v_L$. (d) Estimated launch efficiencies for different launch heights for various lattice beam powers (beam waist $w$=~0.5~mm). (e) Estimated launch efficiencies (height = 1~m) as a function of single lattice beam power and waist.}
\label{fig:lattice}
\end{figure*}

We consider bosonic Cd atoms in the ground state and consider the two-level system made with the $^1$P$_1$ level. In this approximation the optical dipole potential $U\left(\textbf{r}\right)$ can be calculated according to~\cite{Grimm_2000},
\begin{equation}\label{eq:dipolePotential}
    U\left(\textbf{r}\right)=\frac{3\pi c^2}{2\omega_0^3}\left(\frac{\Gamma}{\omega_0-\omega}+\frac{\Gamma}{\omega_0+\omega}\right)I\left(\textbf{r}\right),
\end{equation}
where $\omega$ and $I\left(\textbf{r}\right)$ are, respectively, the angular frequency and intensity profile of the trap light, and $\Gamma$ and $\omega_0$ are the natural linewidth and angular frequency of the two-level system, respectively, in this case the $^1$S$_0$-$^1$P$_1$ transition. 

As is clear from Eq.~\ref{eq:dipolePotential}, Cd is not especially suited to trapping in this manner due to the reduced trap depth coming from the necessarily large values of $\omega_0-\omega$ and especially $\omega_0^3$. Nevertheless, for a optical dipole trap formed from two focused beams at 1064~nm crossing at an angle of 60$^\circ$ in the horizontal plane, trap depths in excess of 30~$\mu$K can be achieved with reasonable powers and waists (Fig.~\ref{fig:lattice}~(a) and~(b)). Commercial lasers at this wavelength can readily produce powers $>$40~W and with M$^2<$1.1.  This means that efficient loading into a dipole trap from the 3D MOT ($T\sim\mu$K) will be available. Due to the lack of known cold collisional properties of Cd, however, it remains to be seen which isotopes or isotope mixture will be suitable for evaporative cooling. In any case, the optical dipole trap can also serve as the initial stage in transferring the prepared atoms $\sim$40~cm into the science chamber of a dual-species interferometer~\cite{Tinsley_2022}. This transfer can also be performed using a 1064~nm laser, with a single shifting focus beam using an optically compensated zoom lens~\cite{Lee_2020}, for which similar waists and powers are needed.

\begin{table}[t]
 \caption{Capture efficiency and loading rates of the 3D MOT at 326 nm for natural $^{114}$Cd and an oven temperature of 100~$^\circ$C. These simulations use 10$^4$ atoms and the parameters given in Table~\ref{tab:beamProperties}. Efficiency and loading rates are shown for different cooling configurations -- see text for more details. Errors are from the counting statistics of the simulation.}
 \label{tab:motLoading}
 \begin{center} 
 \begin{tabular}{l c c}
    \hline
    \hline
           & Efficiency  & Loading Rate (atoms/s) \\
    \hline
       2D MOT \& 3D MOT &   0.02~\%         &   6~$\pm$~4~$\times$~10$^5$  \\
       + Transverse Cooling &   0.34~\%       &     1.0~$\pm$~0.2~$\times~$10$^7$          \\  
       + Crossed Slowing &   0.85~\%         & 2.6~$\pm~$0.3~$\times$~10$^7$  \\
       + Zeeman Slower &   4.13~\%         & 12.6~$\pm~$0.6~$\times$~10$^7$  \\
    \hline
    \hline
 \end{tabular}
 \end{center}
\end{table}

The free evolution time $T$ of the atom interferometer can be enhanced by launching the atoms in a fountain configuration. A dual-species launch requires that an accelerating lattice be used with sufficient trap depth for both atom species, if the atoms are to be simultaneously launched with the along the same spatial trajectory and with the same velocity, a key requirement for minimising systematic errors, although the difference in mass of the two species will result in different trajectories following the application of the interferometry beams. A standing wave based on a high-power 1064-nm laser is well-suited to this task and we calculate the expected launch efficiencies for both Cd and Sr (Fig.~\ref{fig:lattice}~(c)). Our model considers losses from both spontaneous single-photon scattering, which is low due to the large detunings and launch times, and the considerably larger losses due to Landau-Zener tunnelling~\cite{Peik_1997}. To estimate the losses due to Landau-Zener tunnelling, we consider the launch as a sequence of Bloch oscillations and first estimate the trap depth using Eq.~\ref{eq:dipolePotential}, considering a retro-reflected standing-wave configuration~\cite{Hu_2019}. This computed trap depths can be used to determine the band gap energies by numerically solving the Schr\"odinger equation~\cite{Kovachy_2010}. For a launch velocity $v_l$, the fraction of surviving atoms $f$ is then given by~\cite{Sugarbaker_2014},
\begin{equation}\label{eq:lossesLZ}
    f=\left(1-\text{exp}\left[-\frac{\pi\Omega^2}{2\alpha}\right]\right)^N,
\end{equation}
where $\hbar\Omega$ is the band gap energy, $\alpha$ is the chirp rate of the lattice frequency, and $N=\frac{mv_{l}}{2\hbar k_{L}}$ is the number of avoided crossings that the atoms pass through and $k_{L}$ is the lattice wavenumber. Eq.~\ref{eq:lossesLZ} shows that large trap depths and slow chirp rates minimise tunnelling losses.

Figures~\ref{fig:lattice}~(d) and~(e) shows the expected losses for both Cd and Sr as a function of the laser beam power for a final launch height of 1~m and with a lattice acceleration distance of 10~cm. For the used waist of 0.5~mm, the Rayleigh length is 0.7~m and so larger than the launch region. Here the losses from Cd are clearly considerable unless high powers are used due to the reduced depth from the 1/$\omega^3$ dependence of Eq.~\ref{eq:dipolePotential}. This is a more serious issue than for the crossed optical dipole trap, though the required powers for a launch of 1~m with reasonable levels of atom losses (e.g. $<$~50\%) are still experimentally feasible ($\sim$15~W).

The above considerations have assumed that the launch velocity is continuous, whereas in reality it is quantised according to the $2\hbar k_{L}$ momentum imparted by the Bloch oscillations of the lattice undergoing an acceleration $a_{L}$ ~\cite{Peik_1997,Kovachy_2010}. There will therefore be a small difference in the launch velocity for different Cd and Sr isotopes arising from their mass difference. Furthermore, as the Bloch oscillation period also depends upon the mass ($\tau_B=2\hbar k_{L}/ma_{L}$), different isotopes will not in general undergo an equal number of oscillations during the launch and isotopes will be launched in superpositions of momentum states. However, careful selection of the launch characteristics can help suppress these effects, by selecting oscillations close to the ratio of the masses. For example, when considering the two most abundant isotopes, $^{114}$Cd and $^{88}$Sr, a launch with exactly $N_{\text{Cd}}=$~631 oscillations, $N_{\text{Sr}}\approx$~487. In this case the atoms will be launched to 98~cm and the difference in launch velocities will be $\Delta v_{L}=$~0.2~mm/s and the difference in apogees just 0.08~mm.

\section{Conclusion \& Outlook}\label{sec:summary}
We have presented the design and thorough simulation of a state-of-the-art apparatus for producing ultracold Cd samples. The design is cognisant of the unique challenges of this atomic species, especially the broadband dipole-allowed $^1$S$_0$-$^1$P$_1$ transition at 229~nm. Specifically, we have simulated that it is possible to efficiently load Cd atoms directly into an intercombination-transition MOT starting from an atomic oven, by first using the 229~nm to generate a slow atomic beam, overcoming problems associated with photoionsiation. Such a segmented architecture may be useful for other alkaline-earth-like elements, especially Zn whose relevant transitions are similar to Cd (214~nm, $\Gamma$=2$\pi\times$71~MHz and 308~nm, $\Gamma$=2$\pi\times$4~kHz) and whose laser cooling and trapping is in its infancy~\cite{Buki_2021}.

The design is to be used as a basis for atom interferometry~\cite{Tinsley_2022}, where the intercombination transitions of Cd make it a good candidate for both Bragg interferometry~\cite{delAguila2018} and single-photon clock-transition atom interferometry~\cite{Hu_2017,Chiarotti_2022}. Longer term, the device seems compatible with continuous atom laser systems~\cite{Chen_2022} if differential pumping is introduced on the vertical axis, which is currently left free for interferometry beams. Additionally and conversely, it may be noted that the device can also be operated in a pulsed configuration, with the 229~nm only turned on intermittently for bursts lasting the 5~ms or so required to cool and trap in the 2D MOT, whilst the 3D MOT remains on for the whole duration. This may be beneficial for protecting the vacuum windows and the BBO crystal, for which the damage mechanisms are related to sustained exposure to continuous-wave 229~nm. A similar idea has very recently been shown to be highly effective in preventing degradation in an evacuated optical cavity at 244~nm~\cite{Zhadnov_2023}.

\section{Acknowledgments}
We thank Aidan Arnold and Stefan Truppe for useful discussions, Nicola Grani and Shamaila Manzoor for initial work on the simulations, and Leonardo Salvi for help with the magnetic field calculations. This work has been supported by the European Research Council, Grant No.772126 (TICTOCGRAV). J.N.T acknowledges the support of the Horizon Europe Grant ID 101080164 (UVQuanT).

\bibliography{main}

\begin{thebibliography}{82}%
\makeatletter
\providecommand \@ifxundefined [1]{%
 \@ifx{#1\undefined}
}%
\providecommand \@ifnum [1]{%
 \ifnum #1\expandafter \@firstoftwo
 \else \expandafter \@secondoftwo
 \fi
}%
\providecommand \@ifx [1]{%
 \ifx #1\expandafter \@firstoftwo
 \else \expandafter \@secondoftwo
 \fi
}%
\providecommand \natexlab [1]{#1}%
\providecommand \enquote  [1]{``#1''}%
\providecommand \bibnamefont  [1]{#1}%
\providecommand \bibfnamefont [1]{#1}%
\providecommand \citenamefont [1]{#1}%
\providecommand \href@noop [0]{\@secondoftwo}%
\providecommand \href [0]{\begingroup \@sanitize@url \@href}%
\providecommand \@href[1]{\@@startlink{#1}\@@href}%
\providecommand \@@href[1]{\endgroup#1\@@endlink}%
\providecommand \@sanitize@url [0]{\catcode `\\12\catcode `\$12\catcode
  `\&12\catcode `\#12\catcode `\^12\catcode `\_12\catcode `\%12\relax}%
\providecommand \@@startlink[1]{}%
\providecommand \@@endlink[0]{}%
\providecommand \url  [0]{\begingroup\@sanitize@url \@url }%
\providecommand \@url [1]{\endgroup\@href {#1}{\urlprefix }}%
\providecommand \urlprefix  [0]{URL }%
\providecommand \Eprint [0]{\href }%
\providecommand \doibase [0]{https://doi.org/}%
\providecommand \selectlanguage [0]{\@gobble}%
\providecommand \bibinfo  [0]{\@secondoftwo}%
\providecommand \bibfield  [0]{\@secondoftwo}%
\providecommand \translation [1]{[#1]}%
\providecommand \BibitemOpen [0]{}%
\providecommand \bibitemStop [0]{}%
\providecommand \bibitemNoStop [0]{.\EOS\space}%
\providecommand \EOS [0]{\spacefactor3000\relax}%
\providecommand \BibitemShut  [1]{\csname bibitem#1\endcsname}%
\let\auto@bib@innerbib\@empty
\bibitem [{\citenamefont {Schreck}\ and\ \citenamefont
  {Druten}(2021)}]{Schreck_2021}%
  \BibitemOpen
  \bibfield  {author} {\bibinfo {author} {\bibfnamefont {F.}~\bibnamefont
  {Schreck}}\ and\ \bibinfo {author} {\bibfnamefont {K.~v.}\ \bibnamefont
  {Druten}},\ }\bibfield  {title} {\bibinfo {title} {Laser cooling for quantum
  gases},\ }\href {https://doi.org/10.1038/s41567-021-01379-w} {\bibfield
  {journal} {\bibinfo  {journal} {Nature Physics}\ }\textbf {\bibinfo {volume}
  {17}},\ \bibinfo {pages} {1296} (\bibinfo {year} {2021})}\BibitemShut
  {NoStop}%
\bibitem [{\citenamefont {Poli}\ \emph {et~al.}(2013)\citenamefont {Poli},
  \citenamefont {Oates}, \citenamefont {Gill},\ and\ \citenamefont
  {Tino}}]{Poli_2013}%
  \BibitemOpen
  \bibfield  {author} {\bibinfo {author} {\bibfnamefont {N.}~\bibnamefont
  {Poli}}, \bibinfo {author} {\bibfnamefont {C.~W.}\ \bibnamefont {Oates}},
  \bibinfo {author} {\bibfnamefont {P.}~\bibnamefont {Gill}},\ and\ \bibinfo
  {author} {\bibfnamefont {G.~M.}\ \bibnamefont {Tino}},\ }\bibfield  {title}
  {\bibinfo {title} {{Optical atomic clocks}},\ }\href
  {https://doi.org/10.1393/ncr/i2013-10095-x} {\bibfield  {journal} {\bibinfo
  {journal} {{Rivista del Nuovo Cimento}}\ }\textbf {\bibinfo {volume} {36}},\
  \bibinfo {pages} {555} (\bibinfo {year} {2013})}\BibitemShut {NoStop}%
\bibitem [{\citenamefont {Safronova}\ \emph {et~al.}(2018)\citenamefont
  {Safronova}, \citenamefont {Budker}, \citenamefont {DeMille}, \citenamefont
  {Kimball}, \citenamefont {Derevianko},\ and\ \citenamefont
  {Clark}}]{Safronova_2018}%
  \BibitemOpen
  \bibfield  {author} {\bibinfo {author} {\bibfnamefont {M.~S.}\ \bibnamefont
  {Safronova}}, \bibinfo {author} {\bibfnamefont {D.}~\bibnamefont {Budker}},
  \bibinfo {author} {\bibfnamefont {D.}~\bibnamefont {DeMille}}, \bibinfo
  {author} {\bibfnamefont {D.~F.~J.}\ \bibnamefont {Kimball}}, \bibinfo
  {author} {\bibfnamefont {A.}~\bibnamefont {Derevianko}},\ and\ \bibinfo
  {author} {\bibfnamefont {C.~W.}\ \bibnamefont {Clark}},\ }\bibfield  {title}
  {\bibinfo {title} {Search for new physics with atoms and molecules},\ }\href
  {https://doi.org/10.1103/RevModPhys.90.025008} {\bibfield  {journal}
  {\bibinfo  {journal} {Rev. Mod. Phys.}\ }\textbf {\bibinfo {volume} {90}},\
  \bibinfo {pages} {025008} (\bibinfo {year} {2018})}\BibitemShut {NoStop}%
\bibitem [{\citenamefont {Tino}\ and\ \citenamefont
  {Kasevich}(2014)}]{Varenna_2014}%
  \BibitemOpen
  \bibfield  {author} {\bibinfo {author} {\bibfnamefont {G.~M.}\ \bibnamefont
  {Tino}}\ and\ \bibinfo {author} {\bibfnamefont {M.~A.}\ \bibnamefont
  {Kasevich}},\ }\href@noop {} {\emph {\bibinfo {title} {{Atom Interferometry,
  Proceedings of the International School of Physics “Enrico Fermi,” Course
  CLXXXVIII, Varenna 2013}}}}\ (\bibinfo  {publisher} {Società Italiana di
  Fisica and IOS Press},\ \bibinfo {year} {2014})\BibitemShut {NoStop}%
\bibitem [{\citenamefont {Yamaguchi}\ \emph {et~al.}(2019)\citenamefont
  {Yamaguchi}, \citenamefont {Safronova}, \citenamefont {Gibble},\ and\
  \citenamefont {Katori}}]{Yamaguchi_2019}%
  \BibitemOpen
  \bibfield  {author} {\bibinfo {author} {\bibfnamefont {A.}~\bibnamefont
  {Yamaguchi}}, \bibinfo {author} {\bibfnamefont {M.~S.}\ \bibnamefont
  {Safronova}}, \bibinfo {author} {\bibfnamefont {K.}~\bibnamefont {Gibble}},\
  and\ \bibinfo {author} {\bibfnamefont {H.}~\bibnamefont {Katori}},\
  }\bibfield  {title} {\bibinfo {title} {{Narrow-line Cooling and Determination
  of the Magic Wavelength of Cd}},\ }\href
  {https://doi.org/10.1103/PhysRevLett.123.113201} {\bibfield  {journal}
  {\bibinfo  {journal} {Phys. Rev. Lett.}\ }\textbf {\bibinfo {volume} {123}},\
  \bibinfo {pages} {113201} (\bibinfo {year} {2019})}\BibitemShut {NoStop}%
\bibitem [{\citenamefont {Zhang}\ \emph {et~al.}(2021)\citenamefont {Zhang},
  \citenamefont {Liu}, \citenamefont {Fu}, \citenamefont {Sun}, \citenamefont
  {Xu},\ and\ \citenamefont {Wang}}]{Zhang_2021}%
  \BibitemOpen
  \bibfield  {author} {\bibinfo {author} {\bibfnamefont {Y.}~\bibnamefont
  {Zhang}}, \bibinfo {author} {\bibfnamefont {Q.}~\bibnamefont {Liu}}, \bibinfo
  {author} {\bibfnamefont {X.}~\bibnamefont {Fu}}, \bibinfo {author}
  {\bibfnamefont {J.}~\bibnamefont {Sun}}, \bibinfo {author} {\bibfnamefont
  {Z.}~\bibnamefont {Xu}},\ and\ \bibinfo {author} {\bibfnamefont
  {Y.}~\bibnamefont {Wang}},\ }\bibfield  {title} {\bibinfo {title} {A stable
  deep-ultraviolet laser for laser cooling of mercury atoms},\ }\href
  {https://doi.org/https://doi.org/10.1016/j.optlastec.2021.106956} {\bibfield
  {journal} {\bibinfo  {journal} {Optics \& Laser Technology}\ }\textbf
  {\bibinfo {volume} {139}},\ \bibinfo {pages} {106956} (\bibinfo {year}
  {2021})}\BibitemShut {NoStop}%
\bibitem [{\citenamefont {B\"{u}ki}\ \emph {et~al.}(2021)\citenamefont
  {B\"{u}ki}, \citenamefont {R\"{o}ser},\ and\ \citenamefont
  {Stellmer}}]{Buki_2021}%
  \BibitemOpen
  \bibfield  {author} {\bibinfo {author} {\bibfnamefont {M.}~\bibnamefont
  {B\"{u}ki}}, \bibinfo {author} {\bibfnamefont {D.}~\bibnamefont
  {R\"{o}ser}},\ and\ \bibinfo {author} {\bibfnamefont {S.}~\bibnamefont
  {Stellmer}},\ }\bibfield  {title} {\bibinfo {title} {Frequency-quintupled
  laser at 308 nm for atomic physics applications},\ }\href
  {https://doi.org/10.1364/AO.438793} {\bibfield  {journal} {\bibinfo
  {journal} {Appl. Opt.}\ }\textbf {\bibinfo {volume} {60}},\ \bibinfo {pages}
  {9915} (\bibinfo {year} {2021})}\BibitemShut {NoStop}%
\bibitem [{\citenamefont {Lavigne}\ \emph {et~al.}(2022)\citenamefont
  {Lavigne}, \citenamefont {Groh},\ and\ \citenamefont
  {Stellmer}}]{Lavigne_2022}%
  \BibitemOpen
  \bibfield  {author} {\bibinfo {author} {\bibfnamefont {Q.}~\bibnamefont
  {Lavigne}}, \bibinfo {author} {\bibfnamefont {T.}~\bibnamefont {Groh}},\ and\
  \bibinfo {author} {\bibfnamefont {S.}~\bibnamefont {Stellmer}},\ }\bibfield
  {title} {\bibinfo {title} {Magneto-optical trapping of mercury at high
  phase-space density},\ }\href {https://doi.org/10.1103/PhysRevA.105.033106}
  {\bibfield  {journal} {\bibinfo  {journal} {Phys. Rev. A}\ }\textbf {\bibinfo
  {volume} {105}},\ \bibinfo {pages} {033106} (\bibinfo {year}
  {2022})}\BibitemShut {NoStop}%
\bibitem [{\citenamefont {Tinsley}\ \emph {et~al.}(2022)\citenamefont
  {Tinsley}, \citenamefont {Bandarupally}, \citenamefont {Chiarotti},
  \citenamefont {Manzoor}, \citenamefont {Salvi},\ and\ \citenamefont
  {Poli}}]{Tinsley_2022}%
  \BibitemOpen
  \bibfield  {author} {\bibinfo {author} {\bibfnamefont {J.~N.}\ \bibnamefont
  {Tinsley}}, \bibinfo {author} {\bibfnamefont {S.}~\bibnamefont
  {Bandarupally}}, \bibinfo {author} {\bibfnamefont {M.}~\bibnamefont
  {Chiarotti}}, \bibinfo {author} {\bibfnamefont {S.}~\bibnamefont {Manzoor}},
  \bibinfo {author} {\bibfnamefont {L.}~\bibnamefont {Salvi}},\ and\ \bibinfo
  {author} {\bibfnamefont {N.}~\bibnamefont {Poli}},\ }\bibfield  {title}
  {\bibinfo {title} {{Prospects for a simultaneous atom interferometer with
  ultracold cadmium and strontium for fundamental physics tests}},\ }in\ \href
  {https://doi.org/10.1117/12.2616918} {\emph {\bibinfo {booktitle} {Optical
  and Quantum Sensing and Precision Metrology II}}},\ Vol.\ \bibinfo {volume}
  {12016},\ \bibinfo {editor} {edited by\ \bibinfo {editor} {\bibfnamefont
  {J.}~\bibnamefont {Scheuer}}\ and\ \bibinfo {editor} {\bibfnamefont {S.~M.}\
  \bibnamefont {Shahriar}}},\ \bibinfo {organization} {International Society
  for Optics and Photonics}\ (\bibinfo  {publisher} {SPIE},\ \bibinfo {year}
  {2022})\ pp.\ \bibinfo {pages} {1 -- 16}\BibitemShut {NoStop}%
\bibitem [{\citenamefont {Takamoto}\ \emph {et~al.}(2005)\citenamefont
  {Takamoto}, \citenamefont {Hong}, \citenamefont {Higashi},\ and\
  \citenamefont {Katori}}]{Takamoto_2005}%
  \BibitemOpen
  \bibfield  {author} {\bibinfo {author} {\bibfnamefont {M.}~\bibnamefont
  {Takamoto}}, \bibinfo {author} {\bibfnamefont {F.-L.}\ \bibnamefont {Hong}},
  \bibinfo {author} {\bibfnamefont {R.}~\bibnamefont {Higashi}},\ and\ \bibinfo
  {author} {\bibfnamefont {H.}~\bibnamefont {Katori}},\ }\bibfield  {title}
  {\bibinfo {title} {An optical lattice clock},\ }\href
  {https://doi.org/10.1038/nature03541} {\bibfield  {journal} {\bibinfo
  {journal} {Nature}\ }\textbf {\bibinfo {volume} {435}},\ \bibinfo {pages}
  {321} (\bibinfo {year} {2005})}\BibitemShut {NoStop}%
\bibitem [{\citenamefont {McGrew}\ \emph {et~al.}(2018)\citenamefont {McGrew},
  \citenamefont {Zhang}, \citenamefont {Fasano}, \citenamefont {Sch\"{a}ffer},
  \citenamefont {Beloy}, \citenamefont {Nicolodi}, \citenamefont {Brown},
  \citenamefont {Hinkley}, \citenamefont {Milani}, \citenamefont {Schioppo},
  \citenamefont {Yoon},\ and\ \citenamefont {Ludlow}}]{McGrew_2018}%
  \BibitemOpen
  \bibfield  {author} {\bibinfo {author} {\bibfnamefont {W.~F.}\ \bibnamefont
  {McGrew}}, \bibinfo {author} {\bibfnamefont {X.}~\bibnamefont {Zhang}},
  \bibinfo {author} {\bibfnamefont {R.~J.}\ \bibnamefont {Fasano}}, \bibinfo
  {author} {\bibfnamefont {S.~A.}\ \bibnamefont {Sch\"{a}ffer}}, \bibinfo
  {author} {\bibfnamefont {K.}~\bibnamefont {Beloy}}, \bibinfo {author}
  {\bibfnamefont {D.}~\bibnamefont {Nicolodi}}, \bibinfo {author}
  {\bibfnamefont {R.~C.}\ \bibnamefont {Brown}}, \bibinfo {author}
  {\bibfnamefont {N.}~\bibnamefont {Hinkley}}, \bibinfo {author} {\bibfnamefont
  {G.}~\bibnamefont {Milani}}, \bibinfo {author} {\bibfnamefont
  {M.}~\bibnamefont {Schioppo}}, \bibinfo {author} {\bibfnamefont {T.~H.}\
  \bibnamefont {Yoon}},\ and\ \bibinfo {author} {\bibfnamefont {A.~D.}\
  \bibnamefont {Ludlow}},\ }\bibfield  {title} {\bibinfo {title} {Atomic clock
  performance enabling geodesy below the centimetre level},\ }\href
  {https://doi.org/10.1038/s41586-018-0738-2} {\bibfield  {journal} {\bibinfo
  {journal} {Nature}\ }\textbf {\bibinfo {volume} {564}},\ \bibinfo {pages}
  {87} (\bibinfo {year} {2018})}\BibitemShut {NoStop}%
\bibitem [{\citenamefont {Bothwell}\ \emph {et~al.}(2022)\citenamefont
  {Bothwell}, \citenamefont {Kennedy}, \citenamefont {Aeppli}, \citenamefont
  {Kedar}, \citenamefont {Robinson}, \citenamefont {Oelker}, \citenamefont
  {Staron},\ and\ \citenamefont {Ye}}]{Bothwell_2022}%
  \BibitemOpen
  \bibfield  {author} {\bibinfo {author} {\bibfnamefont {T.}~\bibnamefont
  {Bothwell}}, \bibinfo {author} {\bibfnamefont {C.~J.}\ \bibnamefont
  {Kennedy}}, \bibinfo {author} {\bibfnamefont {A.}~\bibnamefont {Aeppli}},
  \bibinfo {author} {\bibfnamefont {D.}~\bibnamefont {Kedar}}, \bibinfo
  {author} {\bibfnamefont {J.~M.}\ \bibnamefont {Robinson}}, \bibinfo {author}
  {\bibfnamefont {E.}~\bibnamefont {Oelker}}, \bibinfo {author} {\bibfnamefont
  {A.}~\bibnamefont {Staron}},\ and\ \bibinfo {author} {\bibfnamefont
  {J.}~\bibnamefont {Ye}},\ }\bibfield  {title} {\bibinfo {title} {Resolving
  the gravitational redshift across a millimetre-scale atomic sample},\ }\href
  {https://doi.org/10.1038/s41586-021-04349-7} {\bibfield  {journal} {\bibinfo
  {journal} {Nature}\ }\textbf {\bibinfo {volume} {602}},\ \bibinfo {pages}
  {420} (\bibinfo {year} {2022})}\BibitemShut {NoStop}%
\bibitem [{\citenamefont {Zheng}\ \emph {et~al.}(2022)\citenamefont {Zheng},
  \citenamefont {Dolde}, \citenamefont {Lochab}, \citenamefont {Merriman},
  \citenamefont {Li},\ and\ \citenamefont {Kolkowitz}}]{Zheng_2022}%
  \BibitemOpen
  \bibfield  {author} {\bibinfo {author} {\bibfnamefont {X.}~\bibnamefont
  {Zheng}}, \bibinfo {author} {\bibfnamefont {J.}~\bibnamefont {Dolde}},
  \bibinfo {author} {\bibfnamefont {V.}~\bibnamefont {Lochab}}, \bibinfo
  {author} {\bibfnamefont {B.~N.}\ \bibnamefont {Merriman}}, \bibinfo {author}
  {\bibfnamefont {H.}~\bibnamefont {Li}},\ and\ \bibinfo {author}
  {\bibfnamefont {S.}~\bibnamefont {Kolkowitz}},\ }\bibfield  {title} {\bibinfo
  {title} {Differential clock comparisons with a multiplexed optical lattice
  clock},\ }\href {https://doi.org/10.1038/s41586-021-04344-y} {\bibfield
  {journal} {\bibinfo  {journal} {Nature}\ }\textbf {\bibinfo {volume} {602}},\
  \bibinfo {pages} {425} (\bibinfo {year} {2022})}\BibitemShut {NoStop}%
\bibitem [{\citenamefont {Hu}\ \emph {et~al.}(2017)\citenamefont {Hu},
  \citenamefont {Poli}, \citenamefont {Salvi},\ and\ \citenamefont
  {Tino}}]{Hu_2017}%
  \BibitemOpen
  \bibfield  {author} {\bibinfo {author} {\bibfnamefont {L.}~\bibnamefont
  {Hu}}, \bibinfo {author} {\bibfnamefont {N.}~\bibnamefont {Poli}}, \bibinfo
  {author} {\bibfnamefont {L.}~\bibnamefont {Salvi}},\ and\ \bibinfo {author}
  {\bibfnamefont {G.~M.}\ \bibnamefont {Tino}},\ }\bibfield  {title} {\bibinfo
  {title} {{Atom Interferometry with the Sr Optical Clock Transition}},\ }\href
  {https://doi.org/10.1103/PhysRevLett.119.263601} {\bibfield  {journal}
  {\bibinfo  {journal} {Phys. Rev. Lett.}\ }\textbf {\bibinfo {volume} {119}},\
  \bibinfo {pages} {263601} (\bibinfo {year} {2017})}\BibitemShut {NoStop}%
\bibitem [{\citenamefont {Badurina}\ \emph {et~al.}(2020)\citenamefont
  {Badurina} \emph {et~al.}}]{AION_2020}%
  \BibitemOpen
  \bibfield  {author} {\bibinfo {author} {\bibfnamefont {L.}~\bibnamefont
  {Badurina}} \emph {et~al.},\ }\bibfield  {title} {\bibinfo {title} {{AION}:
  an atom interferometer observatory and network},\ }\href
  {https://doi.org/10.1088/1475-7516/2020/05/011} {\bibfield  {journal}
  {\bibinfo  {journal} {Journal of Cosmology and Astroparticle Physics}\
  }\textbf {\bibinfo {volume} {2020}}\bibinfo  {number} { (05)},\ \bibinfo
  {pages} {011}}\BibitemShut {NoStop}%
\bibitem [{\citenamefont {Itano}\ \emph {et~al.}(1982)\citenamefont {Itano},
  \citenamefont {Lewis},\ and\ \citenamefont {Wineland}}]{Itano_1982}%
  \BibitemOpen
\bibfield  {number} {  }\bibfield  {author} {\bibinfo {author} {\bibfnamefont
  {W.~M.}\ \bibnamefont {Itano}}, \bibinfo {author} {\bibfnamefont {L.~L.}\
  \bibnamefont {Lewis}},\ and\ \bibinfo {author} {\bibfnamefont {D.~J.}\
  \bibnamefont {Wineland}},\ }\bibfield  {title} {\bibinfo {title} {{Shift of
  $^{2}$S$_{\frac{1}{2}}$ hyperfine splittings due to blackbody radiation}},\
  }\href {https://doi.org/10.1103/PhysRevA.25.1233} {\bibfield  {journal}
  {\bibinfo  {journal} {Phys. Rev. A}\ }\textbf {\bibinfo {volume} {25}},\
  \bibinfo {pages} {1233} (\bibinfo {year} {1982})}\BibitemShut {NoStop}%
\bibitem [{\citenamefont {Bothwell}\ \emph {et~al.}(2019)\citenamefont
  {Bothwell}, \citenamefont {Kedar}, \citenamefont {Oelker}, \citenamefont
  {Robinson}, \citenamefont {Bromley}, \citenamefont {Tew}, \citenamefont
  {Ye},\ and\ \citenamefont {Kennedy}}]{Bothwell_2019}%
  \BibitemOpen
  \bibfield  {author} {\bibinfo {author} {\bibfnamefont {T.}~\bibnamefont
  {Bothwell}}, \bibinfo {author} {\bibfnamefont {D.}~\bibnamefont {Kedar}},
  \bibinfo {author} {\bibfnamefont {E.}~\bibnamefont {Oelker}}, \bibinfo
  {author} {\bibfnamefont {J.~M.}\ \bibnamefont {Robinson}}, \bibinfo {author}
  {\bibfnamefont {S.~L.}\ \bibnamefont {Bromley}}, \bibinfo {author}
  {\bibfnamefont {W.~L.}\ \bibnamefont {Tew}}, \bibinfo {author} {\bibfnamefont
  {J.}~\bibnamefont {Ye}},\ and\ \bibinfo {author} {\bibfnamefont {C.~J.}\
  \bibnamefont {Kennedy}},\ }\bibfield  {title} {\bibinfo {title} {{JILA} {SrI}
  optical lattice clock with uncertainty of $2.0 \times 10^{-18}$},\ }\href
  {https://doi.org/10.1088/1681-7575/ab4089} {\bibfield  {journal} {\bibinfo
  {journal} {Metrologia}\ }\textbf {\bibinfo {volume} {56}},\ \bibinfo {pages}
  {065004} (\bibinfo {year} {2019})}\BibitemShut {NoStop}%
\bibitem [{\citenamefont {Haslinger}\ \emph {et~al.}(2018)\citenamefont
  {Haslinger}, \citenamefont {Jaffe}, \citenamefont {Xu}, \citenamefont
  {Schwartz}, \citenamefont {Sonnleitner}, \citenamefont {Ritsch-Marte},
  \citenamefont {Ritsch},\ and\ \citenamefont {M\"{u}ller}}]{Haslinger_2018}%
  \BibitemOpen
  \bibfield  {author} {\bibinfo {author} {\bibfnamefont {P.}~\bibnamefont
  {Haslinger}}, \bibinfo {author} {\bibfnamefont {M.}~\bibnamefont {Jaffe}},
  \bibinfo {author} {\bibfnamefont {V.}~\bibnamefont {Xu}}, \bibinfo {author}
  {\bibfnamefont {O.}~\bibnamefont {Schwartz}}, \bibinfo {author}
  {\bibfnamefont {M.}~\bibnamefont {Sonnleitner}}, \bibinfo {author}
  {\bibfnamefont {M.}~\bibnamefont {Ritsch-Marte}}, \bibinfo {author}
  {\bibfnamefont {H.}~\bibnamefont {Ritsch}},\ and\ \bibinfo {author}
  {\bibfnamefont {H.}~\bibnamefont {M\"{u}ller}},\ }\bibfield  {title}
  {\bibinfo {title} {Attractive force on atoms due to blackbody radiation},\
  }\href {https://doi.org/10.1038/s41567-017-0004-9} {\bibfield  {journal}
  {\bibinfo  {journal} {Nature Physics}\ }\textbf {\bibinfo {volume} {14}},\
  \bibinfo {pages} {257} (\bibinfo {year} {2018})}\BibitemShut {NoStop}%
\bibitem [{\citenamefont {Kaufman}\ and\ \citenamefont
  {Ni}(2021)}]{Kaufman_2021}%
  \BibitemOpen
  \bibfield  {author} {\bibinfo {author} {\bibfnamefont {A.~M.}\ \bibnamefont
  {Kaufman}}\ and\ \bibinfo {author} {\bibfnamefont {K.-K.}\ \bibnamefont
  {Ni}},\ }\bibfield  {title} {\bibinfo {title} {Quantum science with optical
  tweezer arrays of ultracold atoms and molecules},\ }\href
  {https://doi.org/10.1038/s41567-021-01357-2} {\bibfield  {journal} {\bibinfo
  {journal} {Nature Physics}\ }\textbf {\bibinfo {volume} {17}},\ \bibinfo
  {pages} {1324} (\bibinfo {year} {2021})}\BibitemShut {NoStop}%
\bibitem [{\citenamefont {Brickman}\ \emph {et~al.}(2007)\citenamefont
  {Brickman}, \citenamefont {Chang}, \citenamefont {Acton}, \citenamefont
  {Chew}, \citenamefont {Matsukevich}, \citenamefont {Haljan}, \citenamefont
  {Bagnato},\ and\ \citenamefont {Monroe}}]{Brickman_2007}%
  \BibitemOpen
  \bibfield  {author} {\bibinfo {author} {\bibfnamefont {K.-A.}\ \bibnamefont
  {Brickman}}, \bibinfo {author} {\bibfnamefont {M.-S.}\ \bibnamefont {Chang}},
  \bibinfo {author} {\bibfnamefont {M.}~\bibnamefont {Acton}}, \bibinfo
  {author} {\bibfnamefont {A.}~\bibnamefont {Chew}}, \bibinfo {author}
  {\bibfnamefont {D.}~\bibnamefont {Matsukevich}}, \bibinfo {author}
  {\bibfnamefont {P.~C.}\ \bibnamefont {Haljan}}, \bibinfo {author}
  {\bibfnamefont {V.~S.}\ \bibnamefont {Bagnato}},\ and\ \bibinfo {author}
  {\bibfnamefont {C.}~\bibnamefont {Monroe}},\ }\bibfield  {title} {\bibinfo
  {title} {Magneto-optical trapping of cadmium},\ }\href
  {https://doi.org/10.1103/PhysRevA.76.043411} {\bibfield  {journal} {\bibinfo
  {journal} {Phys. Rev. A}\ }\textbf {\bibinfo {volume} {76}},\ \bibinfo
  {pages} {043411} (\bibinfo {year} {2007})}\BibitemShut {NoStop}%
\bibitem [{\citenamefont {Kaneda}\ \emph {et~al.}(2016)\citenamefont {Kaneda},
  \citenamefont {Yarborough}, \citenamefont {Merzlyak}, \citenamefont
  {Yamaguchi}, \citenamefont {Hayashida}, \citenamefont {Ohmae},\ and\
  \citenamefont {Katori}}]{Kaneda_2016}%
  \BibitemOpen
  \bibfield  {author} {\bibinfo {author} {\bibfnamefont {Y.}~\bibnamefont
  {Kaneda}}, \bibinfo {author} {\bibfnamefont {J.~M.}\ \bibnamefont
  {Yarborough}}, \bibinfo {author} {\bibfnamefont {Y.}~\bibnamefont
  {Merzlyak}}, \bibinfo {author} {\bibfnamefont {A.}~\bibnamefont {Yamaguchi}},
  \bibinfo {author} {\bibfnamefont {K.}~\bibnamefont {Hayashida}}, \bibinfo
  {author} {\bibfnamefont {N.}~\bibnamefont {Ohmae}},\ and\ \bibinfo {author}
  {\bibfnamefont {H.}~\bibnamefont {Katori}},\ }\bibfield  {title} {\bibinfo
  {title} {{Continuous-wave, single-frequency 229 nm laser source for laser
  cooling of cadmium atoms}},\ }\href {https://doi.org/10.1364/OL.41.000705}
  {\bibfield  {journal} {\bibinfo  {journal} {Opt. Lett.}\ }\textbf {\bibinfo
  {volume} {41}},\ \bibinfo {pages} {705} (\bibinfo {year} {2016})}\BibitemShut
  {NoStop}%
\bibitem [{\citenamefont {Chen}\ \emph {et~al.}(2019)\citenamefont {Chen},
  \citenamefont {Bennetts}, \citenamefont {Escudero}, \citenamefont
  {Pasquiou},\ and\ \citenamefont {Schreck}}]{Chen_2019}%
  \BibitemOpen
  \bibfield  {author} {\bibinfo {author} {\bibfnamefont {C.-C.}\ \bibnamefont
  {Chen}}, \bibinfo {author} {\bibfnamefont {S.}~\bibnamefont {Bennetts}},
  \bibinfo {author} {\bibfnamefont {R.~G.}\ \bibnamefont {Escudero}}, \bibinfo
  {author} {\bibfnamefont {B.}~\bibnamefont {Pasquiou}},\ and\ \bibinfo
  {author} {\bibfnamefont {F.}~\bibnamefont {Schreck}},\ }\bibfield  {title}
  {\bibinfo {title} {Continuous guided strontium beam with high phase-space
  density},\ }\href {https://doi.org/10.1103/PhysRevApplied.12.044014}
  {\bibfield  {journal} {\bibinfo  {journal} {Phys. Rev. Applied}\ }\textbf
  {\bibinfo {volume} {12}},\ \bibinfo {pages} {044014} (\bibinfo {year}
  {2019})}\BibitemShut {NoStop}%
\bibitem [{\citenamefont {Schussheim}\ and\ \citenamefont
  {Gibble}(2018)}]{Schussheim2018}%
  \BibitemOpen
  \bibfield  {author} {\bibinfo {author} {\bibfnamefont {D.}~\bibnamefont
  {Schussheim}}\ and\ \bibinfo {author} {\bibfnamefont {K.}~\bibnamefont
  {Gibble}},\ }\bibfield  {title} {\bibinfo {title} {Laser system to laser-cool
  and trap cadmium: towards a cadmium optical lattice clock}\ }(\bibinfo
  {publisher} {Optical Society of America},\ \bibinfo {year} {2018})\ p.\
  \bibinfo {pages} {LTh1F.2}\BibitemShut {NoStop}%
\bibitem [{\citenamefont {Ohayon}\ \emph {et~al.}(2022)\citenamefont {Ohayon},
  \citenamefont {Hofs\"ass}, \citenamefont {Padilla-Castillo}, \citenamefont
  {Wright}, \citenamefont {Meijer}, \citenamefont {Truppe}, \citenamefont
  {Gibble},\ and\ \citenamefont {Sahoo}}]{Ohayon_2022}%
  \BibitemOpen
  \bibfield  {author} {\bibinfo {author} {\bibfnamefont {B.}~\bibnamefont
  {Ohayon}}, \bibinfo {author} {\bibfnamefont {S.}~\bibnamefont {Hofs\"ass}},
  \bibinfo {author} {\bibfnamefont {J.~E.}\ \bibnamefont {Padilla-Castillo}},
  \bibinfo {author} {\bibfnamefont {S.~C.}\ \bibnamefont {Wright}}, \bibinfo
  {author} {\bibfnamefont {G.}~\bibnamefont {Meijer}}, \bibinfo {author}
  {\bibfnamefont {S.}~\bibnamefont {Truppe}}, \bibinfo {author} {\bibfnamefont
  {K.}~\bibnamefont {Gibble}},\ and\ \bibinfo {author} {\bibfnamefont {B.~K.}\
  \bibnamefont {Sahoo}},\ }\bibfield  {title} {\bibinfo {title} {Isotope shifts
  in cadmium as a sensitive probe for physics beyond the standard model},\
  }\href {https://doi.org/10.1088/1367-2630/acacbb} {\bibfield  {journal}
  {\bibinfo  {journal} {New Journal of Physics}\ }\textbf {\bibinfo {volume}
  {24}},\ \bibinfo {pages} {123040} (\bibinfo {year} {2022})}\BibitemShut
  {NoStop}%
\bibitem [{\citenamefont {Lamporesi}\ \emph {et~al.}(2013)\citenamefont
  {Lamporesi}, \citenamefont {Donadello}, \citenamefont {Serafini},\ and\
  \citenamefont {Ferrari}}]{Lamporesi_2013}%
  \BibitemOpen
  \bibfield  {author} {\bibinfo {author} {\bibfnamefont {G.}~\bibnamefont
  {Lamporesi}}, \bibinfo {author} {\bibfnamefont {S.}~\bibnamefont
  {Donadello}}, \bibinfo {author} {\bibfnamefont {S.}~\bibnamefont
  {Serafini}},\ and\ \bibinfo {author} {\bibfnamefont {G.}~\bibnamefont
  {Ferrari}},\ }\bibfield  {title} {\bibinfo {title} {Compact high-flux source
  of cold sodium atoms},\ }\href {https://doi.org/10.1063/1.4808375} {\bibfield
   {journal} {\bibinfo  {journal} {Review of Scientific Instruments}\ }\textbf
  {\bibinfo {volume} {84}},\ \bibinfo {pages} {063102} (\bibinfo {year}
  {2013})}\BibitemShut {NoStop}%
\bibitem [{\citenamefont {Bennetts}\ \emph {et~al.}(2017)\citenamefont
  {Bennetts}, \citenamefont {Chen}, \citenamefont {Pasquiou},\ and\
  \citenamefont {Schreck}}]{Bennetts_2017}%
  \BibitemOpen
  \bibfield  {author} {\bibinfo {author} {\bibfnamefont {S.}~\bibnamefont
  {Bennetts}}, \bibinfo {author} {\bibfnamefont {C.-C.}\ \bibnamefont {Chen}},
  \bibinfo {author} {\bibfnamefont {B.}~\bibnamefont {Pasquiou}},\ and\
  \bibinfo {author} {\bibfnamefont {F.}~\bibnamefont {Schreck}},\ }\bibfield
  {title} {\bibinfo {title} {{Steady-State Magneto-Optical Trap with 100-Fold
  Improved Phase-Space Density}},\ }\href
  {https://doi.org/10.1103/PhysRevLett.119.223202} {\bibfield  {journal}
  {\bibinfo  {journal} {Phys. Rev. Lett.}\ }\textbf {\bibinfo {volume} {119}},\
  \bibinfo {pages} {223202} (\bibinfo {year} {2017})}\BibitemShut {NoStop}%
\bibitem [{\citenamefont {Tinsley}\ \emph {et~al.}(2021)\citenamefont
  {Tinsley}, \citenamefont {Bandarupally}, \citenamefont {Penttinen},
  \citenamefont {Manzoor}, \citenamefont {Ranta}, \citenamefont {Salvi},
  \citenamefont {Guina},\ and\ \citenamefont {Poli}}]{Tinsley_2021}%
  \BibitemOpen
  \bibfield  {author} {\bibinfo {author} {\bibfnamefont {J.~N.}\ \bibnamefont
  {Tinsley}}, \bibinfo {author} {\bibfnamefont {S.}~\bibnamefont
  {Bandarupally}}, \bibinfo {author} {\bibfnamefont {J.-P.}\ \bibnamefont
  {Penttinen}}, \bibinfo {author} {\bibfnamefont {S.}~\bibnamefont {Manzoor}},
  \bibinfo {author} {\bibfnamefont {S.}~\bibnamefont {Ranta}}, \bibinfo
  {author} {\bibfnamefont {L.}~\bibnamefont {Salvi}}, \bibinfo {author}
  {\bibfnamefont {M.}~\bibnamefont {Guina}},\ and\ \bibinfo {author}
  {\bibfnamefont {N.}~\bibnamefont {Poli}},\ }\bibfield  {title} {\bibinfo
  {title} {Watt-level blue light for precision spectroscopy, laser cooling and
  trapping of strontium and cadmium atoms},\ }\href
  {https://doi.org/10.1364/OE.429898} {\bibfield  {journal} {\bibinfo
  {journal} {Opt. Express}\ }\textbf {\bibinfo {volume} {29}},\ \bibinfo
  {pages} {25462} (\bibinfo {year} {2021})}\BibitemShut {NoStop}%
\bibitem [{\citenamefont {Manzoor}\ \emph {et~al.}(2022)\citenamefont
  {Manzoor}, \citenamefont {Tinsley}, \citenamefont {Bandarupally},
  \citenamefont {Chiarotti},\ and\ \citenamefont {Poli}}]{Manzoor_2022}%
  \BibitemOpen
  \bibfield  {author} {\bibinfo {author} {\bibfnamefont {S.}~\bibnamefont
  {Manzoor}}, \bibinfo {author} {\bibfnamefont {J.~N.}\ \bibnamefont
  {Tinsley}}, \bibinfo {author} {\bibfnamefont {S.}~\bibnamefont
  {Bandarupally}}, \bibinfo {author} {\bibfnamefont {M.}~\bibnamefont
  {Chiarotti}},\ and\ \bibinfo {author} {\bibfnamefont {N.}~\bibnamefont
  {Poli}},\ }\bibfield  {title} {\bibinfo {title} {{High-power,
  frequency-quadrupled UV laser source resonant with the $^1$S$_0$-$^3$P$_1$
  narrow intercombination transition of cadmium at 326.2~nm}},\ }\href
  {https://doi.org/10.1364/OL.457979} {\bibfield  {journal} {\bibinfo
  {journal} {Opt. Lett.}\ }\textbf {\bibinfo {volume} {47}},\ \bibinfo {pages}
  {2582} (\bibinfo {year} {2022})}\BibitemShut {NoStop}%
\bibitem [{\citenamefont {Ferrari}\ \emph {et~al.}(2006)\citenamefont
  {Ferrari}, \citenamefont {Poli}, \citenamefont {Sorrentino},\ and\
  \citenamefont {Tino}}]{Ferrari_2006}%
  \BibitemOpen
  \bibfield  {author} {\bibinfo {author} {\bibfnamefont {G.}~\bibnamefont
  {Ferrari}}, \bibinfo {author} {\bibfnamefont {N.}~\bibnamefont {Poli}},
  \bibinfo {author} {\bibfnamefont {F.}~\bibnamefont {Sorrentino}},\ and\
  \bibinfo {author} {\bibfnamefont {G.~M.}\ \bibnamefont {Tino}},\ }\bibfield
  {title} {\bibinfo {title} {{Long-Lived Bloch Oscillations with Bosonic Sr
  Atoms and Application to Gravity Measurement at the Micrometer Scale}},\
  }\href {https://doi.org/10.1103/PhysRevLett.97.060402} {\bibfield  {journal}
  {\bibinfo  {journal} {Phys. Rev. Lett.}\ }\textbf {\bibinfo {volume} {97}},\
  \bibinfo {pages} {060402} (\bibinfo {year} {2006})}\BibitemShut {NoStop}%
\bibitem [{\citenamefont {Itano}\ \emph {et~al.}(1993)\citenamefont {Itano},
  \citenamefont {Bergquist}, \citenamefont {Bollinger}, \citenamefont
  {Gilligan}, \citenamefont {Heinzen}, \citenamefont {Moore}, \citenamefont
  {Raizen},\ and\ \citenamefont {Wineland}}]{Itano_1993}%
  \BibitemOpen
  \bibfield  {author} {\bibinfo {author} {\bibfnamefont {W.~M.}\ \bibnamefont
  {Itano}}, \bibinfo {author} {\bibfnamefont {J.~C.}\ \bibnamefont
  {Bergquist}}, \bibinfo {author} {\bibfnamefont {J.~J.}\ \bibnamefont
  {Bollinger}}, \bibinfo {author} {\bibfnamefont {J.~M.}\ \bibnamefont
  {Gilligan}}, \bibinfo {author} {\bibfnamefont {D.~J.}\ \bibnamefont
  {Heinzen}}, \bibinfo {author} {\bibfnamefont {F.~L.}\ \bibnamefont {Moore}},
  \bibinfo {author} {\bibfnamefont {M.~G.}\ \bibnamefont {Raizen}},\ and\
  \bibinfo {author} {\bibfnamefont {D.~J.}\ \bibnamefont {Wineland}},\
  }\bibfield  {title} {\bibinfo {title} {{Quantum projection noise: Population
  fluctuations in two-level systems}},\ }\href
  {https://doi.org/10.1103/PhysRevA.47.3554} {\bibfield  {journal} {\bibinfo
  {journal} {Phys. Rev. A}\ }\textbf {\bibinfo {volume} {47}},\ \bibinfo
  {pages} {3554} (\bibinfo {year} {1993})}\BibitemShut {NoStop}%
\bibitem [{\citenamefont {Sorrentino}\ \emph {et~al.}(2014)\citenamefont
  {Sorrentino}, \citenamefont {Bodart}, \citenamefont {Cacciapuoti},
  \citenamefont {Lien}, \citenamefont {Prevedelli}, \citenamefont {Rosi},
  \citenamefont {Salvi},\ and\ \citenamefont {Tino}}]{Sorrentino_2014}%
  \BibitemOpen
  \bibfield  {author} {\bibinfo {author} {\bibfnamefont {F.}~\bibnamefont
  {Sorrentino}}, \bibinfo {author} {\bibfnamefont {Q.}~\bibnamefont {Bodart}},
  \bibinfo {author} {\bibfnamefont {L.}~\bibnamefont {Cacciapuoti}}, \bibinfo
  {author} {\bibfnamefont {Y.-H.}\ \bibnamefont {Lien}}, \bibinfo {author}
  {\bibfnamefont {M.}~\bibnamefont {Prevedelli}}, \bibinfo {author}
  {\bibfnamefont {G.}~\bibnamefont {Rosi}}, \bibinfo {author} {\bibfnamefont
  {L.}~\bibnamefont {Salvi}},\ and\ \bibinfo {author} {\bibfnamefont {G.~M.}\
  \bibnamefont {Tino}},\ }\bibfield  {title} {\bibinfo {title} {{Sensitivity
  limits of a Raman atom interferometer as a gravity gradiometer}},\ }\href
  {https://doi.org/10.1103/PhysRevA.89.023607} {\bibfield  {journal} {\bibinfo
  {journal} {Phys. Rev. A}\ }\textbf {\bibinfo {volume} {89}},\ \bibinfo
  {pages} {023607} (\bibinfo {year} {2014})}\BibitemShut {NoStop}%
\bibitem [{\citenamefont {Savoie}\ \emph {et~al.}(2018)\citenamefont {Savoie},
  \citenamefont {Altorio}, \citenamefont {Fang}, \citenamefont {Sidorenkov},
  \citenamefont {Geiger},\ and\ \citenamefont {Landragin}}]{Savoie_2018}%
  \BibitemOpen
  \bibfield  {author} {\bibinfo {author} {\bibfnamefont {D.}~\bibnamefont
  {Savoie}}, \bibinfo {author} {\bibfnamefont {M.}~\bibnamefont {Altorio}},
  \bibinfo {author} {\bibfnamefont {B.}~\bibnamefont {Fang}}, \bibinfo {author}
  {\bibfnamefont {L.~A.}\ \bibnamefont {Sidorenkov}}, \bibinfo {author}
  {\bibfnamefont {R.}~\bibnamefont {Geiger}},\ and\ \bibinfo {author}
  {\bibfnamefont {A.}~\bibnamefont {Landragin}},\ }\bibfield  {title} {\bibinfo
  {title} {Interleaved atom interferometry for high-sensitivity inertial
  measurements},\ }\href {https://doi.org/10.1126/sciadv.aau7948} {\bibfield
  {journal} {\bibinfo  {journal} {Science Advances}\ }\textbf {\bibinfo
  {volume} {4}},\ \bibinfo {pages} {eaau7948} (\bibinfo {year}
  {2018})}\BibitemShut {NoStop}%
\bibitem [{\citenamefont {Quessada}\ \emph {et~al.}(2003)\citenamefont
  {Quessada}, \citenamefont {Kovacich}, \citenamefont {Courtillot},
  \citenamefont {Clairon}, \citenamefont {Santarelli},\ and\ \citenamefont
  {Lemonde}}]{Quessada_2003}%
  \BibitemOpen
  \bibfield  {author} {\bibinfo {author} {\bibfnamefont {A.}~\bibnamefont
  {Quessada}}, \bibinfo {author} {\bibfnamefont {R.~P.}\ \bibnamefont
  {Kovacich}}, \bibinfo {author} {\bibfnamefont {I.}~\bibnamefont
  {Courtillot}}, \bibinfo {author} {\bibfnamefont {A.}~\bibnamefont {Clairon}},
  \bibinfo {author} {\bibfnamefont {G.}~\bibnamefont {Santarelli}},\ and\
  \bibinfo {author} {\bibfnamefont {P.}~\bibnamefont {Lemonde}},\ }\bibfield
  {title} {\bibinfo {title} {{The Dick effect for an optical frequency
  standard}},\ }\href {https://doi.org/10.1088/1464-4266/5/2/373} {\bibfield
  {journal} {\bibinfo  {journal} {Journal of Optics B: Quantum and
  Semiclassical Optics}\ }\textbf {\bibinfo {volume} {5}},\ \bibinfo {pages}
  {S150} (\bibinfo {year} {2003})}\BibitemShut {NoStop}%
\bibitem [{\citenamefont {Xu}\ \emph {et~al.}(2004)\citenamefont {Xu},
  \citenamefont {Persson}, \citenamefont {Svanberg}, \citenamefont {Blagoev},
  \citenamefont {Malcheva}, \citenamefont {Pentchev}, \citenamefont
  {Bi\'emont}, \citenamefont {Campos}, \citenamefont {Ortiz},\ and\
  \citenamefont {Mayo}}]{Xu_2004}%
  \BibitemOpen
  \bibfield  {author} {\bibinfo {author} {\bibfnamefont {H.~L.}\ \bibnamefont
  {Xu}}, \bibinfo {author} {\bibfnamefont {A.}~\bibnamefont {Persson}},
  \bibinfo {author} {\bibfnamefont {S.}~\bibnamefont {Svanberg}}, \bibinfo
  {author} {\bibfnamefont {K.}~\bibnamefont {Blagoev}}, \bibinfo {author}
  {\bibfnamefont {G.}~\bibnamefont {Malcheva}}, \bibinfo {author}
  {\bibfnamefont {V.}~\bibnamefont {Pentchev}}, \bibinfo {author}
  {\bibfnamefont {E.}~\bibnamefont {Bi\'emont}}, \bibinfo {author}
  {\bibfnamefont {J.}~\bibnamefont {Campos}}, \bibinfo {author} {\bibfnamefont
  {M.}~\bibnamefont {Ortiz}},\ and\ \bibinfo {author} {\bibfnamefont
  {R.}~\bibnamefont {Mayo}},\ }\bibfield  {title} {\bibinfo {title} {{Radiative
  lifetime and transition probabilities in
  $\mathrm{Cd}\phantom{\rule{0.2em}{0ex}}I$ and
  $\mathrm{Cd}\phantom{\rule{0.2em}{0ex}}II$}},\ }\href
  {https://doi.org/10.1103/PhysRevA.70.042508} {\bibfield  {journal} {\bibinfo
  {journal} {Phys. Rev. A}\ }\textbf {\bibinfo {volume} {70}},\ \bibinfo
  {pages} {042508} (\bibinfo {year} {2004})}\BibitemShut {NoStop}%
\bibitem [{\citenamefont {Shaw}\ \emph {et~al.}(2021)\citenamefont {Shaw},
  \citenamefont {Hannig},\ and\ \citenamefont {McCarron}}]{Shaw_2021}%
  \BibitemOpen
  \bibfield  {author} {\bibinfo {author} {\bibfnamefont {J.~C.}\ \bibnamefont
  {Shaw}}, \bibinfo {author} {\bibfnamefont {S.}~\bibnamefont {Hannig}},\ and\
  \bibinfo {author} {\bibfnamefont {D.~J.}\ \bibnamefont {McCarron}},\
  }\bibfield  {title} {\bibinfo {title} {{Stable 2 W continuous-wave 261.5 nm
  laser for cooling and trapping aluminum monochloride}},\ }\href
  {https://doi.org/10.1364/OE.441741} {\bibfield  {journal} {\bibinfo
  {journal} {Opt. Express}\ }\textbf {\bibinfo {volume} {29}},\ \bibinfo
  {pages} {37140} (\bibinfo {year} {2021})}\BibitemShut {NoStop}%
\bibitem [{\citenamefont {Bhar}\ \emph {et~al.}(2000)\citenamefont {Bhar},
  \citenamefont {Kumbhakar}, \citenamefont {Chatterjee}, \citenamefont
  {Rudra},\ and\ \citenamefont {Nagahori}}]{Bhar_2000}%
  \BibitemOpen
  \bibfield  {author} {\bibinfo {author} {\bibfnamefont {G.}~\bibnamefont
  {Bhar}}, \bibinfo {author} {\bibfnamefont {P.}~\bibnamefont {Kumbhakar}},
  \bibinfo {author} {\bibfnamefont {U.}~\bibnamefont {Chatterjee}}, \bibinfo
  {author} {\bibfnamefont {A.}~\bibnamefont {Rudra}},\ and\ \bibinfo {author}
  {\bibfnamefont {A.}~\bibnamefont {Nagahori}},\ }\bibfield  {title} {\bibinfo
  {title} {{Widely tunable deep ultraviolet generation in CLBO}},\ }\href
  {https://doi.org/https://doi.org/10.1016/S0030-4018(00)00513-7} {\bibfield
  {journal} {\bibinfo  {journal} {Optics Communications}\ }\textbf {\bibinfo
  {volume} {176}},\ \bibinfo {pages} {199} (\bibinfo {year}
  {2000})}\BibitemShut {NoStop}%
\bibitem [{\citenamefont {Takachiho}\ \emph {et~al.}(2014)\citenamefont
  {Takachiho}, \citenamefont {Yoshimura}, \citenamefont {Takahashi},
  \citenamefont {Imade}, \citenamefont {Sasaki},\ and\ \citenamefont
  {Mori}}]{Takachiho_2014}%
  \BibitemOpen
  \bibfield  {author} {\bibinfo {author} {\bibfnamefont {K.}~\bibnamefont
  {Takachiho}}, \bibinfo {author} {\bibfnamefont {M.}~\bibnamefont
  {Yoshimura}}, \bibinfo {author} {\bibfnamefont {Y.}~\bibnamefont
  {Takahashi}}, \bibinfo {author} {\bibfnamefont {M.}~\bibnamefont {Imade}},
  \bibinfo {author} {\bibfnamefont {T.}~\bibnamefont {Sasaki}},\ and\ \bibinfo
  {author} {\bibfnamefont {Y.}~\bibnamefont {Mori}},\ }\bibfield  {title}
  {\bibinfo {title} {{Ultraviolet laser-induced degradation of
  CsLiB$_6$O$_{10}$ and $\beta$-BaB$_2$O$_4$}},\ }\href
  {https://doi.org/10.1364/OME.4.000559} {\bibfield  {journal} {\bibinfo
  {journal} {Opt. Mater. Express}\ }\textbf {\bibinfo {volume} {4}},\ \bibinfo
  {pages} {559} (\bibinfo {year} {2014})}\BibitemShut {NoStop}%
\bibitem [{\citenamefont {Gebert}\ \emph {et~al.}(2014)\citenamefont {Gebert},
  \citenamefont {Frosz}, \citenamefont {Weiss}, \citenamefont {Wan},
  \citenamefont {Ermolov}, \citenamefont {Joly}, \citenamefont {Schmidt},\ and\
  \citenamefont {Russell}}]{Gebert_2014}%
  \BibitemOpen
  \bibfield  {author} {\bibinfo {author} {\bibfnamefont {F.}~\bibnamefont
  {Gebert}}, \bibinfo {author} {\bibfnamefont {M.~H.}\ \bibnamefont {Frosz}},
  \bibinfo {author} {\bibfnamefont {T.}~\bibnamefont {Weiss}}, \bibinfo
  {author} {\bibfnamefont {Y.}~\bibnamefont {Wan}}, \bibinfo {author}
  {\bibfnamefont {A.}~\bibnamefont {Ermolov}}, \bibinfo {author} {\bibfnamefont
  {N.~Y.}\ \bibnamefont {Joly}}, \bibinfo {author} {\bibfnamefont {P.~O.}\
  \bibnamefont {Schmidt}},\ and\ \bibinfo {author} {\bibfnamefont {P.~S.~J.}\
  \bibnamefont {Russell}},\ }\bibfield  {title} {\bibinfo {title} {{Damage-free
  single-mode transmission of deep-UV light in hollow-core PCF}},\ }\href
  {https://doi.org/10.1364/OE.22.015388} {\bibfield  {journal} {\bibinfo
  {journal} {Opt. Express}\ }\textbf {\bibinfo {volume} {22}},\ \bibinfo
  {pages} {15388} (\bibinfo {year} {2014})}\BibitemShut {NoStop}%
\bibitem [{\citenamefont {Marciniak}\ \emph {et~al.}(2017)\citenamefont
  {Marciniak}, \citenamefont {Ball}, \citenamefont {Hung},\ and\ \citenamefont
  {Biercuk}}]{Marciniak_2017}%
  \BibitemOpen
  \bibfield  {author} {\bibinfo {author} {\bibfnamefont {C.~D.}\ \bibnamefont
  {Marciniak}}, \bibinfo {author} {\bibfnamefont {H.~B.}\ \bibnamefont {Ball}},
  \bibinfo {author} {\bibfnamefont {A.~T.-H.}\ \bibnamefont {Hung}},\ and\
  \bibinfo {author} {\bibfnamefont {M.~J.}\ \bibnamefont {Biercuk}},\
  }\bibfield  {title} {\bibinfo {title} {{Towards fully commercial,
  UV-compatible fiber patch cords}},\ }\href
  {https://doi.org/10.1364/OE.25.015643} {\bibfield  {journal} {\bibinfo
  {journal} {Opt. Express}\ }\textbf {\bibinfo {volume} {25}},\ \bibinfo
  {pages} {15643} (\bibinfo {year} {2017})}\BibitemShut {NoStop}%
\bibitem [{\citenamefont {Hollenshead}\ and\ \citenamefont
  {Klebanoff}(2006)}]{Hollenshead_2006}%
  \BibitemOpen
  \bibfield  {author} {\bibinfo {author} {\bibfnamefont {J.}~\bibnamefont
  {Hollenshead}}\ and\ \bibinfo {author} {\bibfnamefont {L.}~\bibnamefont
  {Klebanoff}},\ }\bibfield  {title} {\bibinfo {title} {{Modeling
  radiation-induced carbon contamination of extreme ultraviolet optics}},\
  }\href {https://doi.org/10.1116/1.2140005} {\bibfield  {journal} {\bibinfo
  {journal} {Journal of Vacuum Science \& Technology B: Microelectronics and
  Nanometer Structures Processing, Measurement, and Phenomena}\ }\textbf
  {\bibinfo {volume} {24}},\ \bibinfo {pages} {64} (\bibinfo {year}
  {2006})}\BibitemShut {NoStop}%
\bibitem [{\citenamefont {Gangloff}\ \emph {et~al.}(2015)\citenamefont
  {Gangloff}, \citenamefont {Shi}, \citenamefont {Wu}, \citenamefont
  {Bylinskii}, \citenamefont {Braverman}, \citenamefont {Gutierrez},
  \citenamefont {Nichols}, \citenamefont {Li}, \citenamefont {Aichholz},
  \citenamefont {Cetina}, \citenamefont {Karpa}, \citenamefont
  {Jelenkovi\'{c}}, \citenamefont {Chuang},\ and\ \citenamefont
  {Vuleti\'{c}}}]{Gangloff_2015}%
  \BibitemOpen
  \bibfield  {author} {\bibinfo {author} {\bibfnamefont {D.}~\bibnamefont
  {Gangloff}}, \bibinfo {author} {\bibfnamefont {M.}~\bibnamefont {Shi}},
  \bibinfo {author} {\bibfnamefont {T.}~\bibnamefont {Wu}}, \bibinfo {author}
  {\bibfnamefont {A.}~\bibnamefont {Bylinskii}}, \bibinfo {author}
  {\bibfnamefont {B.}~\bibnamefont {Braverman}}, \bibinfo {author}
  {\bibfnamefont {M.}~\bibnamefont {Gutierrez}}, \bibinfo {author}
  {\bibfnamefont {R.}~\bibnamefont {Nichols}}, \bibinfo {author} {\bibfnamefont
  {J.}~\bibnamefont {Li}}, \bibinfo {author} {\bibfnamefont {K.}~\bibnamefont
  {Aichholz}}, \bibinfo {author} {\bibfnamefont {M.}~\bibnamefont {Cetina}},
  \bibinfo {author} {\bibfnamefont {L.}~\bibnamefont {Karpa}}, \bibinfo
  {author} {\bibfnamefont {B.}~\bibnamefont {Jelenkovi\'{c}}}, \bibinfo
  {author} {\bibfnamefont {I.}~\bibnamefont {Chuang}},\ and\ \bibinfo {author}
  {\bibfnamefont {V.}~\bibnamefont {Vuleti\'{c}}},\ }\bibfield  {title}
  {\bibinfo {title} {{Preventing and reversing vacuum-induced optical losses in
  high-finesse tantalum (V) oxide mirror coatings}},\ }\href
  {https://doi.org/10.1364/OE.23.018014} {\bibfield  {journal} {\bibinfo
  {journal} {Opt. Express}\ }\textbf {\bibinfo {volume} {23}},\ \bibinfo
  {pages} {18014} (\bibinfo {year} {2015})}\BibitemShut {NoStop}%
\bibitem [{\citenamefont {Hubka}\ \emph {et~al.}(2021)\citenamefont {Hubka},
  \citenamefont {Nov\'{a}k}, \citenamefont {Majerov\'{a}}, \citenamefont
  {Green}, \citenamefont {Velpula}, \citenamefont {Boge}, \citenamefont
  {Antipenkov}, \citenamefont {\v{S}obr}, \citenamefont {Kramer}, \citenamefont
  {Majer}, \citenamefont {Naylon}, \citenamefont {Bakule},\ and\ \citenamefont
  {Rus}}]{Hubka_2021}%
  \BibitemOpen
  \bibfield  {author} {\bibinfo {author} {\bibfnamefont {Z.}~\bibnamefont
  {Hubka}}, \bibinfo {author} {\bibfnamefont {J.}~\bibnamefont {Nov\'{a}k}},
  \bibinfo {author} {\bibfnamefont {I.}~\bibnamefont {Majerov\'{a}}}, \bibinfo
  {author} {\bibfnamefont {J.~T.}\ \bibnamefont {Green}}, \bibinfo {author}
  {\bibfnamefont {P.~K.}\ \bibnamefont {Velpula}}, \bibinfo {author}
  {\bibfnamefont {R.}~\bibnamefont {Boge}}, \bibinfo {author} {\bibfnamefont
  {R.}~\bibnamefont {Antipenkov}}, \bibinfo {author} {\bibfnamefont
  {V.}~\bibnamefont {\v{S}obr}}, \bibinfo {author} {\bibfnamefont
  {D.}~\bibnamefont {Kramer}}, \bibinfo {author} {\bibfnamefont
  {K.}~\bibnamefont {Majer}}, \bibinfo {author} {\bibfnamefont {J.~A.}\
  \bibnamefont {Naylon}}, \bibinfo {author} {\bibfnamefont {P.}~\bibnamefont
  {Bakule}},\ and\ \bibinfo {author} {\bibfnamefont {B.}~\bibnamefont {Rus}},\
  }\bibfield  {title} {\bibinfo {title} {Mitigation of laser-induced
  contamination in vacuum in high-repetition-rate high-peak-power laser
  systems},\ }\href {https://doi.org/10.1364/AO.414878} {\bibfield  {journal}
  {\bibinfo  {journal} {Appl. Opt.}\ }\textbf {\bibinfo {volume} {60}},\
  \bibinfo {pages} {533} (\bibinfo {year} {2021})}\BibitemShut {NoStop}%
\bibitem [{\citenamefont {Burkley}\ \emph {et~al.}(2021)\citenamefont
  {Burkley}, \citenamefont {de~Sousa~Borges}, \citenamefont {Ohayon},
  \citenamefont {Golovizin}, \citenamefont {Zhang},\ and\ \citenamefont
  {Crivelli}}]{Burkley_2021}%
  \BibitemOpen
  \bibfield  {author} {\bibinfo {author} {\bibfnamefont {Z.}~\bibnamefont
  {Burkley}}, \bibinfo {author} {\bibfnamefont {L.}~\bibnamefont
  {de~Sousa~Borges}}, \bibinfo {author} {\bibfnamefont {B.}~\bibnamefont
  {Ohayon}}, \bibinfo {author} {\bibfnamefont {A.}~\bibnamefont {Golovizin}},
  \bibinfo {author} {\bibfnamefont {J.}~\bibnamefont {Zhang}},\ and\ \bibinfo
  {author} {\bibfnamefont {P.}~\bibnamefont {Crivelli}},\ }\bibfield  {title}
  {\bibinfo {title} {Stable high power deep-uv enhancement cavity in ultra-high
  vacuum with fluoride coatings},\ }\href {https://doi.org/10.1364/OE.432552}
  {\bibfield  {journal} {\bibinfo  {journal} {Opt. Express}\ }\textbf {\bibinfo
  {volume} {29}},\ \bibinfo {pages} {27450} (\bibinfo {year}
  {2021})}\BibitemShut {NoStop}%
\bibitem [{\citenamefont {Steane}\ \emph {et~al.}(1992)\citenamefont {Steane},
  \citenamefont {Chowdhury},\ and\ \citenamefont {Foot}}]{Steane_1992}%
  \BibitemOpen
  \bibfield  {author} {\bibinfo {author} {\bibfnamefont {A.~M.}\ \bibnamefont
  {Steane}}, \bibinfo {author} {\bibfnamefont {M.}~\bibnamefont {Chowdhury}},\
  and\ \bibinfo {author} {\bibfnamefont {C.~J.}\ \bibnamefont {Foot}},\
  }\bibfield  {title} {\bibinfo {title} {Radiation force in the magneto-optical
  trap},\ }\href {https://doi.org/10.1364/JOSAB.9.002142} {\bibfield  {journal}
  {\bibinfo  {journal} {J. Opt. Soc. Am. B}\ }\textbf {\bibinfo {volume} {9}},\
  \bibinfo {pages} {2142} (\bibinfo {year} {1992})}\BibitemShut {NoStop}%
\bibitem [{\citenamefont {Kuwamoto}\ \emph {et~al.}(1999)\citenamefont
  {Kuwamoto}, \citenamefont {Honda}, \citenamefont {Takahashi},\ and\
  \citenamefont {Yabuzaki}}]{Kuwamoto_1999}%
  \BibitemOpen
  \bibfield  {author} {\bibinfo {author} {\bibfnamefont {T.}~\bibnamefont
  {Kuwamoto}}, \bibinfo {author} {\bibfnamefont {K.}~\bibnamefont {Honda}},
  \bibinfo {author} {\bibfnamefont {Y.}~\bibnamefont {Takahashi}},\ and\
  \bibinfo {author} {\bibfnamefont {T.}~\bibnamefont {Yabuzaki}},\ }\bibfield
  {title} {\bibinfo {title} {{Magneto-optical trapping of Yb atoms using an
  intercombination transition}},\ }\href
  {https://doi.org/10.1103/PhysRevA.60.R745} {\bibfield  {journal} {\bibinfo
  {journal} {Phys. Rev. A}\ }\textbf {\bibinfo {volume} {60}},\ \bibinfo
  {pages} {R745} (\bibinfo {year} {1999})}\BibitemShut {NoStop}%
\bibitem [{\citenamefont {Hannig}\ \emph {et~al.}(2018)\citenamefont {Hannig},
  \citenamefont {Mielke}, \citenamefont {Fenske}, \citenamefont {Misera},
  \citenamefont {Beev}, \citenamefont {Ospelkaus},\ and\ \citenamefont
  {Schmidt}}]{Hannig_2018}%
  \BibitemOpen
  \bibfield  {author} {\bibinfo {author} {\bibfnamefont {S.}~\bibnamefont
  {Hannig}}, \bibinfo {author} {\bibfnamefont {J.}~\bibnamefont {Mielke}},
  \bibinfo {author} {\bibfnamefont {J.~A.}\ \bibnamefont {Fenske}}, \bibinfo
  {author} {\bibfnamefont {M.}~\bibnamefont {Misera}}, \bibinfo {author}
  {\bibfnamefont {N.}~\bibnamefont {Beev}}, \bibinfo {author} {\bibfnamefont
  {C.}~\bibnamefont {Ospelkaus}},\ and\ \bibinfo {author} {\bibfnamefont
  {P.~O.}\ \bibnamefont {Schmidt}},\ }\bibfield  {title} {\bibinfo {title} {A
  highly stable monolithic enhancement cavity for second harmonic generation in
  the ultraviolet},\ }\href {https://doi.org/10.1063/1.5005515} {\bibfield
  {journal} {\bibinfo  {journal} {Review of Scientific Instruments}\ }\textbf
  {\bibinfo {volume} {89}},\ \bibinfo {pages} {013106} (\bibinfo {year}
  {2018})}\BibitemShut {NoStop}%
\bibitem [{\citenamefont {Chiarotti}\ \emph {et~al.}(2022)\citenamefont
  {Chiarotti}, \citenamefont {Tinsley}, \citenamefont {Bandarupally},
  \citenamefont {Manzoor}, \citenamefont {Sacco}, \citenamefont {Salvi},\ and\
  \citenamefont {Poli}}]{Chiarotti_2022}%
  \BibitemOpen
  \bibfield  {author} {\bibinfo {author} {\bibfnamefont {M.}~\bibnamefont
  {Chiarotti}}, \bibinfo {author} {\bibfnamefont {J.~N.}\ \bibnamefont
  {Tinsley}}, \bibinfo {author} {\bibfnamefont {S.}~\bibnamefont
  {Bandarupally}}, \bibinfo {author} {\bibfnamefont {S.}~\bibnamefont
  {Manzoor}}, \bibinfo {author} {\bibfnamefont {M.}~\bibnamefont {Sacco}},
  \bibinfo {author} {\bibfnamefont {L.}~\bibnamefont {Salvi}},\ and\ \bibinfo
  {author} {\bibfnamefont {N.}~\bibnamefont {Poli}},\ }\bibfield  {title}
  {\bibinfo {title} {{Practical Limits for Large-Momentum-Transfer Clock Atom
  Interferometers}},\ }\href {https://doi.org/10.1103/PRXQuantum.3.030348}
  {\bibfield  {journal} {\bibinfo  {journal} {PRX Quantum}\ }\textbf {\bibinfo
  {volume} {3}},\ \bibinfo {pages} {030348} (\bibinfo {year}
  {2022})}\BibitemShut {NoStop}%
\bibitem [{\citenamefont {Guttridge}\ \emph {et~al.}(2016)\citenamefont
  {Guttridge}, \citenamefont {Hopkins}, \citenamefont {Kemp}, \citenamefont
  {Boddy}, \citenamefont {Freytag}, \citenamefont {Jones}, \citenamefont
  {Tarbutt}, \citenamefont {Hinds},\ and\ \citenamefont
  {Cornish}}]{Guttridge_2016}%
  \BibitemOpen
  \bibfield  {author} {\bibinfo {author} {\bibfnamefont {A.}~\bibnamefont
  {Guttridge}}, \bibinfo {author} {\bibfnamefont {S.~A.}\ \bibnamefont
  {Hopkins}}, \bibinfo {author} {\bibfnamefont {S.~L.}\ \bibnamefont {Kemp}},
  \bibinfo {author} {\bibfnamefont {D.}~\bibnamefont {Boddy}}, \bibinfo
  {author} {\bibfnamefont {R.}~\bibnamefont {Freytag}}, \bibinfo {author}
  {\bibfnamefont {M.~P.~A.}\ \bibnamefont {Jones}}, \bibinfo {author}
  {\bibfnamefont {M.~R.}\ \bibnamefont {Tarbutt}}, \bibinfo {author}
  {\bibfnamefont {E.~A.}\ \bibnamefont {Hinds}},\ and\ \bibinfo {author}
  {\bibfnamefont {S.~L.}\ \bibnamefont {Cornish}},\ }\bibfield  {title}
  {\bibinfo {title} {{Direct loading of a large Yb {MOT} on the
  $^1$S$_0\to^3$P$_1$ transition}},\ }\href
  {https://doi.org/10.1088/0953-4075/49/14/145006} {\bibfield  {journal}
  {\bibinfo  {journal} {Journal of Physics B: Atomic, Molecular and Optical
  Physics}\ }\textbf {\bibinfo {volume} {49}},\ \bibinfo {pages} {145006}
  (\bibinfo {year} {2016})}\BibitemShut {NoStop}%
\bibitem [{\citenamefont {Wodey}\ \emph {et~al.}(2021)\citenamefont {Wodey},
  \citenamefont {Rengelink}, \citenamefont {Meiners}, \citenamefont {Rasel},\
  and\ \citenamefont {Schlippert}}]{Wodey_2021}%
  \BibitemOpen
  \bibfield  {author} {\bibinfo {author} {\bibfnamefont {E.}~\bibnamefont
  {Wodey}}, \bibinfo {author} {\bibfnamefont {R.~J.}\ \bibnamefont
  {Rengelink}}, \bibinfo {author} {\bibfnamefont {C.}~\bibnamefont {Meiners}},
  \bibinfo {author} {\bibfnamefont {E.~M.}\ \bibnamefont {Rasel}},\ and\
  \bibinfo {author} {\bibfnamefont {D.}~\bibnamefont {Schlippert}},\ }\bibfield
   {title} {\bibinfo {title} {A robust, high-flux source of laser-cooled
  ytterbium atoms},\ }\href {https://doi.org/10.1088/1361-6455/abd2d1}
  {\bibfield  {journal} {\bibinfo  {journal} {Journal of Physics B: Atomic,
  Molecular and Optical Physics}\ }\textbf {\bibinfo {volume} {54}},\ \bibinfo
  {pages} {035301} (\bibinfo {year} {2021})}\BibitemShut {NoStop}%
\bibitem [{\citenamefont {Lunden}\ \emph {et~al.}(2020)\citenamefont {Lunden},
  \citenamefont {Du}, \citenamefont {Cantara}, \citenamefont {Barral},
  \citenamefont {Jamison},\ and\ \citenamefont {Ketterle}}]{Lunden_2020}%
  \BibitemOpen
  \bibfield  {author} {\bibinfo {author} {\bibfnamefont {W.}~\bibnamefont
  {Lunden}}, \bibinfo {author} {\bibfnamefont {L.}~\bibnamefont {Du}}, \bibinfo
  {author} {\bibfnamefont {M.}~\bibnamefont {Cantara}}, \bibinfo {author}
  {\bibfnamefont {P.}~\bibnamefont {Barral}}, \bibinfo {author} {\bibfnamefont
  {A.~O.}\ \bibnamefont {Jamison}},\ and\ \bibinfo {author} {\bibfnamefont
  {W.}~\bibnamefont {Ketterle}},\ }\bibfield  {title} {\bibinfo {title}
  {{Enhancing the capture velocity of a Dy magneto-optical trap with two-stage
  slowing}},\ }\href {https://doi.org/10.1103/PhysRevA.101.063403} {\bibfield
  {journal} {\bibinfo  {journal} {Phys. Rev. A}\ }\textbf {\bibinfo {volume}
  {101}},\ \bibinfo {pages} {063403} (\bibinfo {year} {2020})}\BibitemShut
  {NoStop}%
\bibitem [{\citenamefont {Seo}\ \emph {et~al.}(2020)\citenamefont {Seo},
  \citenamefont {Chen}, \citenamefont {Chen}, \citenamefont {Yuan},
  \citenamefont {Huang}, \citenamefont {Du},\ and\ \citenamefont
  {Jo}}]{Seo_2020}%
  \BibitemOpen
  \bibfield  {author} {\bibinfo {author} {\bibfnamefont {B.}~\bibnamefont
  {Seo}}, \bibinfo {author} {\bibfnamefont {P.}~\bibnamefont {Chen}}, \bibinfo
  {author} {\bibfnamefont {Z.}~\bibnamefont {Chen}}, \bibinfo {author}
  {\bibfnamefont {W.}~\bibnamefont {Yuan}}, \bibinfo {author} {\bibfnamefont
  {M.}~\bibnamefont {Huang}}, \bibinfo {author} {\bibfnamefont
  {S.}~\bibnamefont {Du}},\ and\ \bibinfo {author} {\bibfnamefont {G.-B.}\
  \bibnamefont {Jo}},\ }\bibfield  {title} {\bibinfo {title} {Efficient
  production of a narrow-line erbium magneto-optical trap with two-stage
  slowing},\ }\href {https://doi.org/10.1103/PhysRevA.102.013319} {\bibfield
  {journal} {\bibinfo  {journal} {Phys. Rev. A}\ }\textbf {\bibinfo {volume}
  {102}},\ \bibinfo {pages} {013319} (\bibinfo {year} {2020})}\BibitemShut
  {NoStop}%
\bibitem [{\citenamefont {Plotkin-Swing}\ \emph {et~al.}(2020)\citenamefont
  {Plotkin-Swing}, \citenamefont {Wirth}, \citenamefont {Gochnauer},
  \citenamefont {Rahman}, \citenamefont {McAlpine},\ and\ \citenamefont
  {Gupta}}]{Plotkin-Swing_2020}%
  \BibitemOpen
  \bibfield  {author} {\bibinfo {author} {\bibfnamefont {B.}~\bibnamefont
  {Plotkin-Swing}}, \bibinfo {author} {\bibfnamefont {A.}~\bibnamefont
  {Wirth}}, \bibinfo {author} {\bibfnamefont {D.}~\bibnamefont {Gochnauer}},
  \bibinfo {author} {\bibfnamefont {T.}~\bibnamefont {Rahman}}, \bibinfo
  {author} {\bibfnamefont {K.~E.}\ \bibnamefont {McAlpine}},\ and\ \bibinfo
  {author} {\bibfnamefont {S.}~\bibnamefont {Gupta}},\ }\bibfield  {title}
  {\bibinfo {title} {Crossed-beam slowing to enhance narrow-line ytterbium
  magneto-optic traps},\ }\href {https://doi.org/10.1063/5.0011361} {\bibfield
  {journal} {\bibinfo  {journal} {Review of Scientific Instruments}\ }\textbf
  {\bibinfo {volume} {91}},\ \bibinfo {pages} {093201} (\bibinfo {year}
  {2020})}\BibitemShut {NoStop}%
\bibitem [{\citenamefont {Wohlleben}\ \emph {et~al.}(2001)\citenamefont
  {Wohlleben}, \citenamefont {Chevy}, \citenamefont {Madison},\ and\
  \citenamefont {Dalibard}}]{Wohlleben_2001}%
  \BibitemOpen
  \bibfield  {author} {\bibinfo {author} {\bibfnamefont {W.}~\bibnamefont
  {Wohlleben}}, \bibinfo {author} {\bibfnamefont {F.}~\bibnamefont {Chevy}},
  \bibinfo {author} {\bibfnamefont {K.}~\bibnamefont {Madison}},\ and\ \bibinfo
  {author} {\bibfnamefont {J.}~\bibnamefont {Dalibard}},\ }\bibfield  {title}
  {\bibinfo {title} {An atom faucet},\ }\href
  {https://doi.org/10.1007/s100530170171} {\bibfield  {journal} {\bibinfo
  {journal} {The European Physical Journal D - Atomic, Molecular, Optical and
  Plasma Physics}\ }\textbf {\bibinfo {volume} {15}},\ \bibinfo {pages} {237}
  (\bibinfo {year} {2001})}\BibitemShut {NoStop}%
\bibitem [{\citenamefont {Chaudhuri}\ \emph {et~al.}(2006)\citenamefont
  {Chaudhuri}, \citenamefont {Roy},\ and\ \citenamefont
  {Unnikrishnan}}]{Chaudhuri_2006}%
  \BibitemOpen
  \bibfield  {author} {\bibinfo {author} {\bibfnamefont {S.}~\bibnamefont
  {Chaudhuri}}, \bibinfo {author} {\bibfnamefont {S.}~\bibnamefont {Roy}},\
  and\ \bibinfo {author} {\bibfnamefont {C.~S.}\ \bibnamefont {Unnikrishnan}},\
  }\bibfield  {title} {\bibinfo {title} {{Realization of an intense cold Rb
  atomic beam based on a two-dimensional magneto-optical trap: Experiments and
  comparison with simulations}},\ }\href
  {https://doi.org/10.1103/PhysRevA.74.023406} {\bibfield  {journal} {\bibinfo
  {journal} {Phys. Rev. A}\ }\textbf {\bibinfo {volume} {74}},\ \bibinfo
  {pages} {023406} (\bibinfo {year} {2006})}\BibitemShut {NoStop}%
\bibitem [{\citenamefont {Barbiero}\ \emph {et~al.}(2020)\citenamefont
  {Barbiero}, \citenamefont {Tarallo}, \citenamefont {Calonico}, \citenamefont
  {Levi}, \citenamefont {Lamporesi},\ and\ \citenamefont
  {Ferrari}}]{Barbiero_2020}%
  \BibitemOpen
  \bibfield  {author} {\bibinfo {author} {\bibfnamefont {M.}~\bibnamefont
  {Barbiero}}, \bibinfo {author} {\bibfnamefont {M.~G.}\ \bibnamefont
  {Tarallo}}, \bibinfo {author} {\bibfnamefont {D.}~\bibnamefont {Calonico}},
  \bibinfo {author} {\bibfnamefont {F.}~\bibnamefont {Levi}}, \bibinfo {author}
  {\bibfnamefont {G.}~\bibnamefont {Lamporesi}},\ and\ \bibinfo {author}
  {\bibfnamefont {G.}~\bibnamefont {Ferrari}},\ }\bibfield  {title} {\bibinfo
  {title} {{Sideband-Enhanced Cold Atomic Source for Optical Clocks}},\ }\href
  {https://doi.org/10.1103/PhysRevApplied.13.014013} {\bibfield  {journal}
  {\bibinfo  {journal} {Phys. Rev. Applied}\ }\textbf {\bibinfo {volume}
  {13}},\ \bibinfo {pages} {014013} (\bibinfo {year} {2020})}\BibitemShut
  {NoStop}%
\bibitem [{\citenamefont {Kohel}\ \emph {et~al.}(2003)\citenamefont {Kohel},
  \citenamefont {Ramirez-Serrano}, \citenamefont {Thompson}, \citenamefont
  {Maleki}, \citenamefont {Bliss},\ and\ \citenamefont
  {Libbrecht}}]{Kohel_2003}%
  \BibitemOpen
  \bibfield  {author} {\bibinfo {author} {\bibfnamefont {J.~M.}\ \bibnamefont
  {Kohel}}, \bibinfo {author} {\bibfnamefont {J.}~\bibnamefont
  {Ramirez-Serrano}}, \bibinfo {author} {\bibfnamefont {R.~J.}\ \bibnamefont
  {Thompson}}, \bibinfo {author} {\bibfnamefont {L.}~\bibnamefont {Maleki}},
  \bibinfo {author} {\bibfnamefont {J.~L.}\ \bibnamefont {Bliss}},\ and\
  \bibinfo {author} {\bibfnamefont {K.~G.}\ \bibnamefont {Libbrecht}},\
  }\bibfield  {title} {\bibinfo {title} {Generation of an intense cold-atom
  beam from a pyramidal magneto-optical trap: experiment and simulation},\
  }\href {https://doi.org/10.1364/JOSAB.20.001161} {\bibfield  {journal}
  {\bibinfo  {journal} {J. Opt. Soc. Am. B}\ }\textbf {\bibinfo {volume}
  {20}},\ \bibinfo {pages} {1161} (\bibinfo {year} {2003})}\BibitemShut
  {NoStop}%
\bibitem [{\citenamefont {Bounds}\ \emph {et~al.}(2018)\citenamefont {Bounds},
  \citenamefont {Jackson}, \citenamefont {Hanley}, \citenamefont {Faoro},
  \citenamefont {Bridge}, \citenamefont {Huillery},\ and\ \citenamefont
  {Jones}}]{Bounds_2018}%
  \BibitemOpen
  \bibfield  {author} {\bibinfo {author} {\bibfnamefont {A.~D.}\ \bibnamefont
  {Bounds}}, \bibinfo {author} {\bibfnamefont {N.~C.}\ \bibnamefont {Jackson}},
  \bibinfo {author} {\bibfnamefont {R.~K.}\ \bibnamefont {Hanley}}, \bibinfo
  {author} {\bibfnamefont {R.}~\bibnamefont {Faoro}}, \bibinfo {author}
  {\bibfnamefont {E.~M.}\ \bibnamefont {Bridge}}, \bibinfo {author}
  {\bibfnamefont {P.}~\bibnamefont {Huillery}},\ and\ \bibinfo {author}
  {\bibfnamefont {M.~P.~A.}\ \bibnamefont {Jones}},\ }\bibfield  {title}
  {\bibinfo {title} {{Rydberg-Dressed Magneto-optical Trap}},\ }\href
  {https://doi.org/10.1103/PhysRevLett.120.183401} {\bibfield  {journal}
  {\bibinfo  {journal} {Phys. Rev. Lett.}\ }\textbf {\bibinfo {volume} {120}},\
  \bibinfo {pages} {183401} (\bibinfo {year} {2018})}\BibitemShut {NoStop}%
\bibitem [{\citenamefont {Hanley}\ \emph {et~al.}(2018)\citenamefont {Hanley},
  \citenamefont {Huillery}, \citenamefont {Keegan}, \citenamefont {Bounds},
  \citenamefont {Boddy}, \citenamefont {Faoro},\ and\ \citenamefont
  {Jones}}]{Hanley_2018}%
  \BibitemOpen
  \bibfield  {author} {\bibinfo {author} {\bibfnamefont {R.~K.}\ \bibnamefont
  {Hanley}}, \bibinfo {author} {\bibfnamefont {P.}~\bibnamefont {Huillery}},
  \bibinfo {author} {\bibfnamefont {N.~C.}\ \bibnamefont {Keegan}}, \bibinfo
  {author} {\bibfnamefont {A.~D.}\ \bibnamefont {Bounds}}, \bibinfo {author}
  {\bibfnamefont {D.}~\bibnamefont {Boddy}}, \bibinfo {author} {\bibfnamefont
  {R.}~\bibnamefont {Faoro}},\ and\ \bibinfo {author} {\bibfnamefont
  {M.~P.~A.}\ \bibnamefont {Jones}},\ }\bibfield  {title} {\bibinfo {title}
  {Quantitative simulation of a magneto-optical trap operating near the photon
  recoil limit},\ }\href {https://doi.org/10.1080/09500340.2017.1401679}
  {\bibfield  {journal} {\bibinfo  {journal} {Journal of Modern Optics}\
  }\textbf {\bibinfo {volume} {65}},\ \bibinfo {pages} {667} (\bibinfo {year}
  {2018})}\BibitemShut {NoStop}%
\bibitem [{\citenamefont {Bondi}(1964)}]{Bondi_1964}%
  \BibitemOpen
  \bibfield  {author} {\bibinfo {author} {\bibfnamefont {A.}~\bibnamefont
  {Bondi}},\ }\bibfield  {title} {\bibinfo {title} {{van der Waals Volumes and
  Radii}},\ }\href {https://doi.org/10.1021/j100785a001} {\bibfield  {journal}
  {\bibinfo  {journal} {The Journal of Physical Chemistry}\ }\textbf {\bibinfo
  {volume} {68}},\ \bibinfo {pages} {441} (\bibinfo {year} {1964})}\BibitemShut
  {NoStop}%
\bibitem [{\citenamefont {Schioppo}\ \emph {et~al.}(2012)\citenamefont
  {Schioppo}, \citenamefont {Poli}, \citenamefont {Prevedelli}, \citenamefont
  {Falke}, \citenamefont {Lisdat}, \citenamefont {Sterr},\ and\ \citenamefont
  {Tino}}]{Schioppo_2012}%
  \BibitemOpen
  \bibfield  {author} {\bibinfo {author} {\bibfnamefont {M.}~\bibnamefont
  {Schioppo}}, \bibinfo {author} {\bibfnamefont {N.}~\bibnamefont {Poli}},
  \bibinfo {author} {\bibfnamefont {M.}~\bibnamefont {Prevedelli}}, \bibinfo
  {author} {\bibfnamefont {S.}~\bibnamefont {Falke}}, \bibinfo {author}
  {\bibfnamefont {C.}~\bibnamefont {Lisdat}}, \bibinfo {author} {\bibfnamefont
  {U.}~\bibnamefont {Sterr}},\ and\ \bibinfo {author} {\bibfnamefont {G.~M.}\
  \bibnamefont {Tino}},\ }\bibfield  {title} {\bibinfo {title} {A compact and
  efficient strontium oven for laser-cooling experiments},\ }\href
  {https://doi.org/10.1063/1.4756936} {\bibfield  {journal} {\bibinfo
  {journal} {Review of Scientific Instruments}\ }\textbf {\bibinfo {volume}
  {83}},\ \bibinfo {pages} {103101} (\bibinfo {year} {2012})}\BibitemShut
  {NoStop}%
\bibitem [{\citenamefont {Gao}\ \emph {et~al.}(2014)\citenamefont {Gao},
  \citenamefont {Liu}, \citenamefont {Xu}, \citenamefont {Tian}, \citenamefont
  {Wang}, \citenamefont {Ren}, \citenamefont {Wu},\ and\ \citenamefont
  {Chang}}]{Gao_2014}%
  \BibitemOpen
  \bibfield  {author} {\bibinfo {author} {\bibfnamefont {F.}~\bibnamefont
  {Gao}}, \bibinfo {author} {\bibfnamefont {H.}~\bibnamefont {Liu}}, \bibinfo
  {author} {\bibfnamefont {P.}~\bibnamefont {Xu}}, \bibinfo {author}
  {\bibfnamefont {X.}~\bibnamefont {Tian}}, \bibinfo {author} {\bibfnamefont
  {Y.}~\bibnamefont {Wang}}, \bibinfo {author} {\bibfnamefont {J.}~\bibnamefont
  {Ren}}, \bibinfo {author} {\bibfnamefont {H.}~\bibnamefont {Wu}},\ and\
  \bibinfo {author} {\bibfnamefont {H.}~\bibnamefont {Chang}},\ }\bibfield
  {title} {\bibinfo {title} {Precision measurement of transverse velocity
  distribution of a strontium atomic beam},\ }\href
  {https://doi.org/10.1063/1.4866983} {\bibfield  {journal} {\bibinfo
  {journal} {AIP Advances}\ }\textbf {\bibinfo {volume} {4}},\ \bibinfo {pages}
  {027118} (\bibinfo {year} {2014})}\BibitemShut {NoStop}%
\bibitem [{\citenamefont {Greenland}\ \emph {et~al.}(1985)\citenamefont
  {Greenland}, \citenamefont {Lauder},\ and\ \citenamefont
  {Wort}}]{Greenland_1985}%
  \BibitemOpen
  \bibfield  {author} {\bibinfo {author} {\bibfnamefont {P.~T.}\ \bibnamefont
  {Greenland}}, \bibinfo {author} {\bibfnamefont {M.~A.}\ \bibnamefont
  {Lauder}},\ and\ \bibinfo {author} {\bibfnamefont {D.~J.~H.}\ \bibnamefont
  {Wort}},\ }\bibfield  {title} {\bibinfo {title} {Atomic beam velocity
  distributions},\ }\href {https://doi.org/10.1088/0022-3727/18/7/009}
  {\bibfield  {journal} {\bibinfo  {journal} {Journal of Physics D: Applied
  Physics}\ }\textbf {\bibinfo {volume} {18}},\ \bibinfo {pages} {1223}
  (\bibinfo {year} {1985})}\BibitemShut {NoStop}%
\bibitem [{\citenamefont {Vangeleyn}\ \emph {et~al.}(2009)\citenamefont
  {Vangeleyn}, \citenamefont {Griffin}, \citenamefont {Riis},\ and\
  \citenamefont {Arnold}}]{Vangeleyn_2009}%
  \BibitemOpen
  \bibfield  {author} {\bibinfo {author} {\bibfnamefont {M.}~\bibnamefont
  {Vangeleyn}}, \bibinfo {author} {\bibfnamefont {P.~F.}\ \bibnamefont
  {Griffin}}, \bibinfo {author} {\bibfnamefont {E.}~\bibnamefont {Riis}},\ and\
  \bibinfo {author} {\bibfnamefont {A.~S.}\ \bibnamefont {Arnold}},\ }\bibfield
   {title} {\bibinfo {title} {Single-laser, one beam, tetrahedral
  magneto-optical trap},\ }\href {https://doi.org/10.1364/OE.17.013601}
  {\bibfield  {journal} {\bibinfo  {journal} {Opt. Express}\ }\textbf {\bibinfo
  {volume} {17}},\ \bibinfo {pages} {13601} (\bibinfo {year}
  {2009})}\BibitemShut {NoStop}%
\bibitem [{\citenamefont {Ovchinnikov}(2005)}]{Ovchinnikov_2005}%
  \BibitemOpen
  \bibfield  {author} {\bibinfo {author} {\bibfnamefont {Y.~B.}\ \bibnamefont
  {Ovchinnikov}},\ }\bibfield  {title} {\bibinfo {title} {Compact
  magneto-optical sources of slow atoms},\ }\href
  {https://doi.org/https://doi.org/10.1016/j.optcom.2005.01.047} {\bibfield
  {journal} {\bibinfo  {journal} {Optics Communications}\ }\textbf {\bibinfo
  {volume} {249}},\ \bibinfo {pages} {473} (\bibinfo {year}
  {2005})}\BibitemShut {NoStop}%
\bibitem [{\citenamefont {{Letokhov}}\ \emph {et~al.}(1977)\citenamefont
  {{Letokhov}}, \citenamefont {{Minogin}},\ and\ \citenamefont
  {{Pavlik}}}]{Letokhov_1977}%
  \BibitemOpen
  \bibfield  {author} {\bibinfo {author} {\bibfnamefont {V.~S.}\ \bibnamefont
  {{Letokhov}}}, \bibinfo {author} {\bibfnamefont {V.~G.}\ \bibnamefont
  {{Minogin}}},\ and\ \bibinfo {author} {\bibfnamefont {B.~D.}\ \bibnamefont
  {{Pavlik}}},\ }\bibfield  {title} {\bibinfo {title} {{Cooling and capture of
  atoms and molecules by a resonant light field}},\ }\href@noop {} {\bibfield
  {journal} {\bibinfo  {journal} {Soviet Journal of Experimental and
  Theoretical Physics}\ }\textbf {\bibinfo {volume} {45}},\ \bibinfo {pages}
  {698} (\bibinfo {year} {1977})}\BibitemShut {NoStop}%
\bibitem [{\citenamefont {Cheiney}\ \emph {et~al.}(2011)\citenamefont
  {Cheiney}, \citenamefont {Carraz}, \citenamefont {Bartoszek-Bober},
  \citenamefont {Faure}, \citenamefont {Vermersch}, \citenamefont {Fabre},
  \citenamefont {Gattobigio}, \citenamefont {Lahaye}, \citenamefont
  {Gu\'{e}ry-Odelin},\ and\ \citenamefont {Mathevet}}]{Cheiney_2011}%
  \BibitemOpen
  \bibfield  {author} {\bibinfo {author} {\bibfnamefont {P.}~\bibnamefont
  {Cheiney}}, \bibinfo {author} {\bibfnamefont {O.}~\bibnamefont {Carraz}},
  \bibinfo {author} {\bibfnamefont {D.}~\bibnamefont {Bartoszek-Bober}},
  \bibinfo {author} {\bibfnamefont {S.}~\bibnamefont {Faure}}, \bibinfo
  {author} {\bibfnamefont {F.}~\bibnamefont {Vermersch}}, \bibinfo {author}
  {\bibfnamefont {C.~M.}\ \bibnamefont {Fabre}}, \bibinfo {author}
  {\bibfnamefont {G.~L.}\ \bibnamefont {Gattobigio}}, \bibinfo {author}
  {\bibfnamefont {T.}~\bibnamefont {Lahaye}}, \bibinfo {author} {\bibfnamefont
  {D.}~\bibnamefont {Gu\'{e}ry-Odelin}},\ and\ \bibinfo {author} {\bibfnamefont
  {R.}~\bibnamefont {Mathevet}},\ }\bibfield  {title} {\bibinfo {title} {{A
  Zeeman slower design with permanent magnets in a Halbach configuration}},\
  }\href {https://doi.org/10.1063/1.3600897} {\bibfield  {journal} {\bibinfo
  {journal} {Review of Scientific Instruments}\ }\textbf {\bibinfo {volume}
  {82}},\ \bibinfo {pages} {063115} (\bibinfo {year} {2011})}\BibitemShut
  {NoStop}%
\bibitem [{\citenamefont {Ali}\ \emph {et~al.}(2017)\citenamefont {Ali},
  \citenamefont {Badr}, \citenamefont {Br{\'{e}}zillon}, \citenamefont
  {Dubessy}, \citenamefont {Perrin},\ and\ \citenamefont {Perrin}}]{Ali_2017}%
  \BibitemOpen
  \bibfield  {author} {\bibinfo {author} {\bibfnamefont {D.~B.}\ \bibnamefont
  {Ali}}, \bibinfo {author} {\bibfnamefont {T.}~\bibnamefont {Badr}}, \bibinfo
  {author} {\bibfnamefont {T.}~\bibnamefont {Br{\'{e}}zillon}}, \bibinfo
  {author} {\bibfnamefont {R.}~\bibnamefont {Dubessy}}, \bibinfo {author}
  {\bibfnamefont {H.}~\bibnamefont {Perrin}},\ and\ \bibinfo {author}
  {\bibfnamefont {A.}~\bibnamefont {Perrin}},\ }\bibfield  {title} {\bibinfo
  {title} {{Detailed study of a transverse field Zeeman slower}},\ }\href
  {https://doi.org/10.1088/1361-6455/aa5a6a} {\bibfield  {journal} {\bibinfo
  {journal} {Journal of Physics B: Atomic, Molecular and Optical Physics}\
  }\textbf {\bibinfo {volume} {50}},\ \bibinfo {pages} {055008} (\bibinfo
  {year} {2017})}\BibitemShut {NoStop}%
\bibitem [{\citenamefont {Ovchinnikov}(2007)}]{Ovchinnikov_2007}%
  \BibitemOpen
  \bibfield  {author} {\bibinfo {author} {\bibfnamefont {Y.~B.}\ \bibnamefont
  {Ovchinnikov}},\ }\bibfield  {title} {\bibinfo {title} {{A Zeeman slower
  based on magnetic dipoles}},\ }\href
  {https://doi.org/https://doi.org/10.1016/j.optcom.2007.04.048} {\bibfield
  {journal} {\bibinfo  {journal} {Optics Communications}\ }\textbf {\bibinfo
  {volume} {276}},\ \bibinfo {pages} {261} (\bibinfo {year}
  {2007})}\BibitemShut {NoStop}%
\bibitem [{\citenamefont {Hill}\ \emph {et~al.}(2014)\citenamefont {Hill},
  \citenamefont {Ovchinnikov}, \citenamefont {Bridge}, \citenamefont {Curtis},\
  and\ \citenamefont {Gill}}]{Hill_2014}%
  \BibitemOpen
  \bibfield  {author} {\bibinfo {author} {\bibfnamefont {I.~R.}\ \bibnamefont
  {Hill}}, \bibinfo {author} {\bibfnamefont {Y.~B.}\ \bibnamefont
  {Ovchinnikov}}, \bibinfo {author} {\bibfnamefont {E.~M.}\ \bibnamefont
  {Bridge}}, \bibinfo {author} {\bibfnamefont {E.~A.}\ \bibnamefont {Curtis}},\
  and\ \bibinfo {author} {\bibfnamefont {P.}~\bibnamefont {Gill}},\ }\bibfield
  {title} {\bibinfo {title} {Zeeman slowers for strontium based on permanent
  magnets},\ }\href {https://doi.org/10.1088/0953-4075/47/7/075006} {\bibfield
  {journal} {\bibinfo  {journal} {Journal of Physics B: Atomic, Molecular and
  Optical Physics}\ }\textbf {\bibinfo {volume} {47}},\ \bibinfo {pages}
  {075006} (\bibinfo {year} {2014})}\BibitemShut {NoStop}%
\bibitem [{\citenamefont {Berglund}\ and\ \citenamefont
  {Wieser}(2011)}]{Berglund_2011}%
  \BibitemOpen
  \bibfield  {author} {\bibinfo {author} {\bibfnamefont {M.}~\bibnamefont
  {Berglund}}\ and\ \bibinfo {author} {\bibfnamefont {M.~E.}\ \bibnamefont
  {Wieser}},\ }\bibfield  {title} {\bibinfo {title} {{Isotopic compositions of
  the elements 2009 (IUPAC Technical Report)}},\ }\href
  {https://doi.org/doi:10.1351/PAC-REP-10-06-02} {\bibfield  {journal}
  {\bibinfo  {journal} {Pure and Applied Chemistry}\ }\textbf {\bibinfo
  {volume} {83}},\ \bibinfo {pages} {397} (\bibinfo {year} {2011})}\BibitemShut
  {NoStop}%
\bibitem [{\citenamefont {Hofs\"ass}\ \emph {et~al.}(2023)\citenamefont
  {Hofs\"ass}, \citenamefont {Padilla-Castillo}, \citenamefont {Wright},
  \citenamefont {Kray}, \citenamefont {Thomas}, \citenamefont {Sartakov},
  \citenamefont {Ohayon}, \citenamefont {Meijer},\ and\ \citenamefont
  {Truppe}}]{Hofsass_2023}%
  \BibitemOpen
  \bibfield  {author} {\bibinfo {author} {\bibfnamefont {S.}~\bibnamefont
  {Hofs\"ass}}, \bibinfo {author} {\bibfnamefont {J.~E.}\ \bibnamefont
  {Padilla-Castillo}}, \bibinfo {author} {\bibfnamefont {S.~C.}\ \bibnamefont
  {Wright}}, \bibinfo {author} {\bibfnamefont {S.}~\bibnamefont {Kray}},
  \bibinfo {author} {\bibfnamefont {R.}~\bibnamefont {Thomas}}, \bibinfo
  {author} {\bibfnamefont {B.~G.}\ \bibnamefont {Sartakov}}, \bibinfo {author}
  {\bibfnamefont {B.}~\bibnamefont {Ohayon}}, \bibinfo {author} {\bibfnamefont
  {G.}~\bibnamefont {Meijer}},\ and\ \bibinfo {author} {\bibfnamefont
  {S.}~\bibnamefont {Truppe}},\ }\bibfield  {title} {\bibinfo {title}
  {{High-resolution isotope-shift spectroscopy of Cd I}},\ }\href
  {https://doi.org/10.1103/PhysRevResearch.5.013043} {\bibfield  {journal}
  {\bibinfo  {journal} {Phys. Rev. Res.}\ }\textbf {\bibinfo {volume} {5}},\
  \bibinfo {pages} {013043} (\bibinfo {year} {2023})}\BibitemShut {NoStop}%
\bibitem [{\citenamefont {Adams}\ \emph {et~al.}(1995)\citenamefont {Adams},
  \citenamefont {Lee}, \citenamefont {Davidson}, \citenamefont {Kasevich},\
  and\ \citenamefont {Chu}}]{Adams_1995}%
  \BibitemOpen
  \bibfield  {author} {\bibinfo {author} {\bibfnamefont {C.~S.}\ \bibnamefont
  {Adams}}, \bibinfo {author} {\bibfnamefont {H.~J.}\ \bibnamefont {Lee}},
  \bibinfo {author} {\bibfnamefont {N.}~\bibnamefont {Davidson}}, \bibinfo
  {author} {\bibfnamefont {M.}~\bibnamefont {Kasevich}},\ and\ \bibinfo
  {author} {\bibfnamefont {S.}~\bibnamefont {Chu}},\ }\bibfield  {title}
  {\bibinfo {title} {{Evaporative Cooling in a Crossed Dipole Trap}},\ }\href
  {https://doi.org/10.1103/PhysRevLett.74.3577} {\bibfield  {journal} {\bibinfo
   {journal} {Phys. Rev. Lett.}\ }\textbf {\bibinfo {volume} {74}},\ \bibinfo
  {pages} {3577} (\bibinfo {year} {1995})}\BibitemShut {NoStop}%
\bibitem [{\citenamefont {Barrett}\ \emph {et~al.}(2001)\citenamefont
  {Barrett}, \citenamefont {Sauer},\ and\ \citenamefont
  {Chapman}}]{Barrett_2001}%
  \BibitemOpen
  \bibfield  {author} {\bibinfo {author} {\bibfnamefont {M.~D.}\ \bibnamefont
  {Barrett}}, \bibinfo {author} {\bibfnamefont {J.~A.}\ \bibnamefont {Sauer}},\
  and\ \bibinfo {author} {\bibfnamefont {M.~S.}\ \bibnamefont {Chapman}},\
  }\bibfield  {title} {\bibinfo {title} {{All-Optical Formation of an Atomic
  Bose-Einstein Condensate}},\ }\href
  {https://doi.org/10.1103/PhysRevLett.87.010404} {\bibfield  {journal}
  {\bibinfo  {journal} {Phys. Rev. Lett.}\ }\textbf {\bibinfo {volume} {87}},\
  \bibinfo {pages} {010404} (\bibinfo {year} {2001})}\BibitemShut {NoStop}%
\bibitem [{\citenamefont {Grimm}\ \emph {et~al.}(2000)\citenamefont {Grimm},
  \citenamefont {Weidemüller},\ and\ \citenamefont
  {Ovchinnikov}}]{Grimm_2000}%
  \BibitemOpen
  \bibfield  {author} {\bibinfo {author} {\bibfnamefont {R.}~\bibnamefont
  {Grimm}}, \bibinfo {author} {\bibfnamefont {M.}~\bibnamefont
  {Weidemüller}},\ and\ \bibinfo {author} {\bibfnamefont {Y.~B.}\ \bibnamefont
  {Ovchinnikov}},\ }\bibfield  {title} {\bibinfo {title} {{Optical Dipole Traps
  for Neutral Atoms}}\ }(\bibinfo  {publisher} {Academic Press},\ \bibinfo
  {year} {2000})\ pp.\ \bibinfo {pages} {95--170}\BibitemShut {NoStop}%
\bibitem [{\citenamefont {Lee}\ \emph {et~al.}(2020)\citenamefont {Lee},
  \citenamefont {Jung}, \citenamefont {Choi},\ and\ \citenamefont
  {Mun}}]{Lee_2020}%
  \BibitemOpen
  \bibfield  {author} {\bibinfo {author} {\bibfnamefont {J.~H.}\ \bibnamefont
  {Lee}}, \bibinfo {author} {\bibfnamefont {H.}~\bibnamefont {Jung}}, \bibinfo
  {author} {\bibfnamefont {J.-y.}\ \bibnamefont {Choi}},\ and\ \bibinfo
  {author} {\bibfnamefont {J.}~\bibnamefont {Mun}},\ }\bibfield  {title}
  {\bibinfo {title} {Transporting cold atoms using an optically compensated
  zoom lens},\ }\href {https://doi.org/10.1103/PhysRevA.102.063106} {\bibfield
  {journal} {\bibinfo  {journal} {Phys. Rev. A}\ }\textbf {\bibinfo {volume}
  {102}},\ \bibinfo {pages} {063106} (\bibinfo {year} {2020})}\BibitemShut
  {NoStop}%
\bibitem [{\citenamefont {Peik}\ \emph {et~al.}(1997)\citenamefont {Peik},
  \citenamefont {Ben~Dahan}, \citenamefont {Bouchoule}, \citenamefont
  {Castin},\ and\ \citenamefont {Salomon}}]{Peik_1997}%
  \BibitemOpen
  \bibfield  {author} {\bibinfo {author} {\bibfnamefont {E.}~\bibnamefont
  {Peik}}, \bibinfo {author} {\bibfnamefont {M.}~\bibnamefont {Ben~Dahan}},
  \bibinfo {author} {\bibfnamefont {I.}~\bibnamefont {Bouchoule}}, \bibinfo
  {author} {\bibfnamefont {Y.}~\bibnamefont {Castin}},\ and\ \bibinfo {author}
  {\bibfnamefont {C.}~\bibnamefont {Salomon}},\ }\bibfield  {title} {\bibinfo
  {title} {Bloch oscillations of atoms, adiabatic rapid passage, and
  monokinetic atomic beams},\ }\href {https://doi.org/10.1103/PhysRevA.55.2989}
  {\bibfield  {journal} {\bibinfo  {journal} {Phys. Rev. A}\ }\textbf {\bibinfo
  {volume} {55}},\ \bibinfo {pages} {2989} (\bibinfo {year}
  {1997})}\BibitemShut {NoStop}%
\bibitem [{\citenamefont {Hu}\ \emph {et~al.}(2020)\citenamefont {Hu},
  \citenamefont {Wang}, \citenamefont {Salvi}, \citenamefont {Tinsley},
  \citenamefont {Tino},\ and\ \citenamefont {Poli}}]{Hu_2019}%
  \BibitemOpen
  \bibfield  {author} {\bibinfo {author} {\bibfnamefont {L.}~\bibnamefont
  {Hu}}, \bibinfo {author} {\bibfnamefont {E.}~\bibnamefont {Wang}}, \bibinfo
  {author} {\bibfnamefont {L.}~\bibnamefont {Salvi}}, \bibinfo {author}
  {\bibfnamefont {J.~N.}\ \bibnamefont {Tinsley}}, \bibinfo {author}
  {\bibfnamefont {G.~M.}\ \bibnamefont {Tino}},\ and\ \bibinfo {author}
  {\bibfnamefont {N.}~\bibnamefont {Poli}},\ }\bibfield  {title} {\bibinfo
  {title} {Sr atom interferometry with the optical clock transition as a
  gravimeter and a gravity gradiometer},\ }\href
  {https://doi.org/10.1088/1361-6382/ab4d18} {\bibfield  {journal} {\bibinfo
  {journal} {Classical and Quantum Gravity}\ }\textbf {\bibinfo {volume}
  {37}},\ \bibinfo {pages} {014001} (\bibinfo {year} {2020})}\BibitemShut
  {NoStop}%
\bibitem [{\citenamefont {Kovachy}\ \emph {et~al.}(2010)\citenamefont
  {Kovachy}, \citenamefont {Hogan}, \citenamefont {Johnson},\ and\
  \citenamefont {Kasevich}}]{Kovachy_2010}%
  \BibitemOpen
  \bibfield  {author} {\bibinfo {author} {\bibfnamefont {T.}~\bibnamefont
  {Kovachy}}, \bibinfo {author} {\bibfnamefont {J.~M.}\ \bibnamefont {Hogan}},
  \bibinfo {author} {\bibfnamefont {D.~M.~S.}\ \bibnamefont {Johnson}},\ and\
  \bibinfo {author} {\bibfnamefont {M.~A.}\ \bibnamefont {Kasevich}},\
  }\bibfield  {title} {\bibinfo {title} {{Optical lattices as waveguides and
  beam splitters for atom interferometry: An analytical treatment and proposal
  of applications}},\ }\href {https://doi.org/10.1103/PhysRevA.82.013638}
  {\bibfield  {journal} {\bibinfo  {journal} {Phys. Rev. A}\ }\textbf {\bibinfo
  {volume} {82}},\ \bibinfo {pages} {013638} (\bibinfo {year}
  {2010})}\BibitemShut {NoStop}%
\bibitem [{\citenamefont {Sugarbaker}(2014)}]{Sugarbaker_2014}%
  \BibitemOpen
  \bibfield  {author} {\bibinfo {author} {\bibfnamefont {A.}~\bibnamefont
  {Sugarbaker}},\ }\emph {\bibinfo {title} {{Atom Interferometry in a 10 m
  Fountain}}},\ \href@noop {} {Ph.D. thesis},\ \bibinfo  {school} {Stanford
  University} (\bibinfo {year} {2014})\BibitemShut {NoStop}%
\bibitem [{\citenamefont {del Aguila}\ \emph {et~al.}(2018)\citenamefont {del
  Aguila}, \citenamefont {Mazzoni}, \citenamefont {Hu}, \citenamefont {Salvi},
  \citenamefont {Tino},\ and\ \citenamefont {Poli}}]{delAguila2018}%
  \BibitemOpen
  \bibfield  {author} {\bibinfo {author} {\bibfnamefont {R.~P.}\ \bibnamefont
  {del Aguila}}, \bibinfo {author} {\bibfnamefont {T.}~\bibnamefont {Mazzoni}},
  \bibinfo {author} {\bibfnamefont {L.}~\bibnamefont {Hu}}, \bibinfo {author}
  {\bibfnamefont {L.}~\bibnamefont {Salvi}}, \bibinfo {author} {\bibfnamefont
  {G.~M.}\ \bibnamefont {Tino}},\ and\ \bibinfo {author} {\bibfnamefont
  {N.}~\bibnamefont {Poli}},\ }\bibfield  {title} {\bibinfo {title} {{Bragg
  gravity-gradiometer using the $^1S_0$-${}^3P_1$ intercombination transition
  of $^{88}$Sr}},\ }\href {https://doi.org/10.1088/1367-2630/aab088} {\bibfield
   {journal} {\bibinfo  {journal} {New J. Phys.}\ }\textbf {\bibinfo {volume}
  {20}},\ \bibinfo {pages} {043002} (\bibinfo {year} {2018})}\BibitemShut
  {NoStop}%
\bibitem [{\citenamefont {Chen}\ \emph {et~al.}(2022)\citenamefont {Chen},
  \citenamefont {Gonz\'alez~Escudero}, \citenamefont {Min\'a\v{r}},
  \citenamefont {Pasquiou}, \citenamefont {Bennetts},\ and\ \citenamefont
  {Schreck}}]{Chen_2022}%
  \BibitemOpen
  \bibfield  {author} {\bibinfo {author} {\bibfnamefont {C.-C.}\ \bibnamefont
  {Chen}}, \bibinfo {author} {\bibfnamefont {R.}~\bibnamefont
  {Gonz\'alez~Escudero}}, \bibinfo {author} {\bibfnamefont {J.}~\bibnamefont
  {Min\'a\v{r}}}, \bibinfo {author} {\bibfnamefont {B.}~\bibnamefont
  {Pasquiou}}, \bibinfo {author} {\bibfnamefont {S.}~\bibnamefont {Bennetts}},\
  and\ \bibinfo {author} {\bibfnamefont {F.}~\bibnamefont {Schreck}},\
  }\bibfield  {title} {\bibinfo {title} {{Continuous Bose–Einstein
  condensation}},\ }\href {https://doi.org/10.1038/s41586-022-04731-z}
  {\bibfield  {journal} {\bibinfo  {journal} {Nature}\ }\textbf {\bibinfo
  {volume} {606}},\ \bibinfo {pages} {683} (\bibinfo {year}
  {2022})}\BibitemShut {NoStop}%
\bibitem [{\citenamefont {Zhadnov}\ \emph {et~al.}(2023)\citenamefont
  {Zhadnov}, \citenamefont {Golovizin}, \citenamefont {Cortinovis},
  \citenamefont {Ohayon}, \citenamefont {de~Sousa~Borges}, \citenamefont
  {Janka},\ and\ \citenamefont {Crivelli}}]{Zhadnov_2023}%
  \BibitemOpen
  \bibfield  {author} {\bibinfo {author} {\bibfnamefont {N.}~\bibnamefont
  {Zhadnov}}, \bibinfo {author} {\bibfnamefont {A.}~\bibnamefont {Golovizin}},
  \bibinfo {author} {\bibfnamefont {I.}~\bibnamefont {Cortinovis}}, \bibinfo
  {author} {\bibfnamefont {B.}~\bibnamefont {Ohayon}}, \bibinfo {author}
  {\bibfnamefont {L.}~\bibnamefont {de~Sousa~Borges}}, \bibinfo {author}
  {\bibfnamefont {G.}~\bibnamefont {Janka}},\ and\ \bibinfo {author}
  {\bibfnamefont {P.}~\bibnamefont {Crivelli}},\ }\href@noop {} {\bibinfo
  {title} {{Pulsed CW laser for long-term spectroscopic measurements at high
  power in deep-UV}}} (\bibinfo {year} {2023}),\ \Eprint
  {https://arxiv.org/abs/2304.13527} {arXiv:2304.13527 [physics.atom-ph]}
  \BibitemShut {NoStop}%
\end{thebibliography}%

\end{document}